\documentclass[lettersize,journal]{IEEEtran}
\usepackage{amsmath,amsfonts}
\usepackage{algorithmic}
\usepackage{array}
\usepackage{textcomp}
\usepackage{stfloats}
\usepackage{url}
\usepackage{verbatim}
\usepackage{graphicx}
\usepackage{cite}
\usepackage{amssymb}
\usepackage{color,colortbl}
\usepackage{pgfplots}
\usetikzlibrary{calc}
\usepackage{multirow}
\usepackage{circuitikz}
\usepackage{subcaption}

\usepackage{tikz}
\usepackage{environ}
\makeatletter
\newsavebox{\measure@tikzpicture}
\NewEnviron{scaletikzpicturetowidth}[1]{%
  \def\tikz@width{#1}%
  \begin{lrbox}{\measure@tikzpicture}%
  \BODY
  \end{lrbox}%
  \pgfmathparse{#1/\wd\measure@tikzpicture}%
  \BODY
}
\usepackage[linesnumbered,ruled,lined]{algorithm2e}

\usepackage[colorlinks=true, allcolors=blue]{hyperref}

\begin{document}

\title{Unit Cell Phase-Frequency Profile Optimization in RIS-Assisted Wide-Band OFDM Systems }

\author{Omran Abbas,~\IEEEmembership{Student~Member,~IEEE,}
Qurrat-Ul-Ain Nadeem,~\IEEEmembership{Member,~IEEE,}
Lo\"{i}c~Markley,~\IEEEmembership{Senior Member,~IEEE,}
and Anas Chaaban,~\IEEEmembership{Senior Member,~IEEE}
        % <-this % stops a space
\thanks{The authors are with the School of Engineering, University of British Columbia, Kelowna, Canada (e-mail: omran.abbas, qurrat.nadeem, loic.markley, anas.chaaban@ubc.ca).}}% <-this % stops a space


\maketitle

\begin{abstract} 
The reflection characteristics of a reconfigurable intelligent surface (RIS) depend on the phase response of the constituent unit cells, which is necessarily frequency dependent. This paper investigates the role of an RIS constituting unit cells with different phase-frequency profiles in a wide-band orthogonal frequency division multiplexing (OFDM) system to improve the achievable rate. We first propose a mathematical model for the phase-frequency relationship parametrized by the phase-frequency profile's slope and phase-shift corresponding to a realizable resonant RIS unit cell. Then, modelling each RIS element with $b$ control bits,  we propose a method for selecting the parameter pairs to obtain a set of $2^b$ phase-frequency profiles. The proposed method yields an RIS design that outperforms existing designs over a wide range of user locations in a single-input, single-output (SISO) OFDM system. We then propose a low-complexity optimization algorithm to maximize the data rate through the joint optimization of (a) power allocations across the sub-carriers and (b) phase-frequency profile for each RIS unit cell from the available set. The analysis is then extended to a multi-user multiple-input single-output (MISO) OFDM scenario. Numerical results show an improvement in the coverage and achievable rates under the proposed framework as compared to single-slope phase-frequency profiles.
\end{abstract}
\vspace{-.1in}

\vspace{-.1in}

\section{Introduction}
Next-generation wireless communication networks face a significant challenge in meeting the high data rate demands of a growing number of devices connecting to them. The key technologies proposed in the last decade to meet these requirements include millimetre wave (mmWave) communication systems that exploit the large bandwidths available at these frequencies \cite{mmwave}, and massive multiple-input multiple-output (MIMO) systems that exploit spatial multiplexing and beamforming gains offered by large antenna arrays  \cite{MIMOCom}. Various other solutions like base station clustering \cite{Cluster}, user-centric ultra-dense networks \cite{usercentric}, non-orthogonal multiple access (NOMA) \cite{noma}, and advanced signal detection methods \cite{machinedense} have also been proposed. However, many of these solutions have high power consumption, cost, signal processing, and computational requirements. An alternative approach to improve the quality of service (QoS) deploys relays as intermediate nodes between the base station (BS) and user equipment (UE) \cite{Relay}. However, relay deployment also faces challenges related to power consumption and hardware cost. 

%There is another type of RIS where a power amplifier is added to the unit cell to increase the strength of reflected signals \cite{amplifierRIS}; this comes at the expense of additional complexity in the unit cell design increased power consumption. This study focuses on power-efficient RIS so the following discussions will be on the former type of RIS.

The deployment of reconfigurable intelligent surfaces (RISs) on different structures in the environment has recently emerged as an alternative approach to enhance the performance of wireless communication systems by controlling the propagation of radio waves between the BS and UEs \cite{RISoverview}. An RIS is an array of reflecting elements, that acts as a spatial filter to alter the characteristics of the incident electromagnetic wave by introducing changes in phase, amplitude \cite{RISPro}, and/or polarization \cite{RISPolar}. The reconfigurability of the RIS elements is typically achieved by embedding low-cost, low-power active elements (e.g., varactor diodes, PIN diodes) in the surface. By changing the biasing state of these elements, the response of the RIS elements can be controlled adaptively. Therefore RISs have the potential to enhance coverage, increase data rates, and enable communication when direct links between the transmitter and receiver are blocked  \cite{THzBlockedLoS}, all the while offering advantages of low power consumption and deployment cost.

%RISs are particularly useful when the link between the transmitter and the receiver is blocked (i.e. no line-of-sight), which is a situation that will become more and more common as future networks will operate at higher frequencies that are more susceptible to blockages. 

Numerous works have incorporated RISs in multiple-input single-output (MISO) systems \cite{MISORIS}, multiple-input multiple-output (MIMO) systems \cite{MIMORIS}, NOMA systems \cite{NOMARIS}, and orthogonal frequency division multiplexing (OFDM) systems \cite{IRSOFDM}. However, all these studies assume that the amplitude and phase responses of the RIS are ideal, that is each element of the RIS exhibits a unity reflection amplitude and introduces a phase shift that is either constant over frequency or can be independently controlled at different frequencies. However, the practical realization of an RIS with elements that exhibit flat or independent phase responses is infeasible in all but narrowband applications. In practice, the phase response of an RIS element is typically very frequency dependent, dropping with frequency over the operation band as the element undergoes resonance. The variation of the RIS response with frequency can sometimes be neglected in a narrowband system operating close to the carrier frequency but cannot be ignored in systems operating over wider bands. Therefore, while the ideal phase-frequency profile assumption can provide accurate data-rate predictions for narrow-band systems, it leads to discrepancies between theoretical and practical results for wide-band systems, where the difference between the practical and ideal phase-frequency profiles becomes more prominent. Accurate characterization of reflection losses and phase-frequency profiles is crucial to minimize such discrepancies.

To address this issue, a few works have studied the effect of the variation in the RIS amplitude and phase response with frequency on the system performance. In \cite{RISstableband}, the authors study different practical  RIS models and introduce the metrics of bandwidth and area of influence to quantify the frequency and geographical ranges respectively, where the RIS can significantly improve the system performance. The authors show that an RIS  has a more significant dynamically adjustable influence on the wave reflection inside a given frequency range. In \cite{PracPin}, the authors studied an RIS-assisted OFDM system considering a frequency-dependent model for the RIS elements' reflection coefficients. The authors concluded that the data rate is improved when using high resolution of discrete phase shifts per each RIS element, but the hardware complexity increases due to the large number of PIN diodes needed per element. In \cite{PracPhaseAmp2} a closed-form equation capturing the phase-dependent amplitude variation in the reflection coefficient of each RIS element is provided. The authors then applied this model to an RIS-assisted narrowband single-UE MISO  communication system and concluded that RIS beamforming based on an idealized reflection amplitude model may lead to significant performance loss as compared to a more practical model. %However they did not incorporate the dependence of the reflection coefficient on the frequency of the incident wave in the proposed model and analysis.
In \cite{PracPhaseAmp}, the authors provided an accurate approximation of the reflection phase and amplitude variation over frequency for a practical  RIS design. The authors then utilized the developed amplitude- and phase-frequency profiles to optimize an RIS-assisted wide-band multi-UE MISO OFDM system by jointly designing the transmit beamforming and RIS reflection coefficients to maximize the average sum rate over all sub-carriers. The phase-frequency profile proposed in \cite{PracPhaseAmp} makes the RIS design and optimization problem complex and more challenging to solve compared with when a flat phase-frequency profile is assumed. Moreover, this work mainly assumes varying zero-crossing frequency of the phase-frequency profiles. However, in this work, we study the case where both the slope and zero-crossing frequency of the phase-frequency profile is controllable.

To address the limitations of existing works, our goals in this paper are to (i) develop a simple and tractable mathematical model that describes the phase-frequency profile of each unit cell in a practical RIS design, (ii) select a set of unit cell phase-frequency profiles that yield good coverage in terms of achievable rate in a desired region, and (iii) optimize the RIS phase shifts to maximize the achievable rate given the phase-frequency profiles set selected above.

%It is worth mentioning that the previous works on RIS-assisted wireless communication systems assume perfect channel state information (CSI) at the base station (BS), e.g., \cite{PracPhaseAmp}, \cite{PracAmpLowComplex}. Acquiring perfect CSI is a challenging task in practice, especially in the presence of the RIS, since channel estimation becomes a time-consuming problem which makes the system unrealistic; for instance, in \cite{CSIOne},  CSI is obtained by turning on and off the RIS elements; which is time-consuming and requires significant overhead, and is has low efficiency in the high mobility areas. Grouping the RIS elements and applying a specific pattern for each group is done in \cite{CSIML}, leading to an efficient algorithm for CSI while deploying large RIS.

We first propose a realistic mathematical model for the phase-frequency profile of each unit cell in  a practical RIS design based on full-wave electromagnetic simulations. The design considers an array of resonant elements printed over a ground plane. The results show that the reflection coefficient's amplitude and phase vary with the incident wave's frequency. By choosing the shape of the unit cell, we can tailor the frequency and quality factor of the resonance to achieve a suitable phase-frequency profile. We show that the phase response of a unit cell can be approximated by an arctangent function parameterized by slope and shift parameters. In order to restrict the RIS unit cells to a discrete set of reconfigurable states, the slopes and shift parameters of the phase-frequency model are constrained to take values from the discrete sets denoted as $\mathcal{M}$ and $\mathcal{I}$ respectively. Based on  the proposed phase-frequency profile  model, we study an RIS-assisted wide-band  OFDM  communication system and optimize its  response to maximize the achievable rate. Our main contributions are summarized as follows:

%make two main contributions (1) optimize the RIS design in terms of the sets developed for shift and slope parameters and (2) optimize the RIS phase shifts based on the proposed model and sets to maximize the achievable rate at the UE in a single-UE SISO system, and (3)  as well as the RIS , where the main focus is the effect of the practical phase profile of the RIS unit cells. 
\begin{itemize}
    \item We analyze the phase-frequency profiles of the reflection coefficient of a unit cell for a practical RIS model. Based on the analysis, we provide a simple mathematical model to approximate the phase-frequency profile over a wide frequency band, that is parameterized by the slope and shift of the phase response. We constrain these parameters to the discrete values specified in the sets $\mathcal{M}$ and $\mathcal{I}$.

    \item We outline the SISO OFDM system model, and obtain a closed-form expression for the optimal RIS phase-frequency profile for each unit cell that would maximize the achievable rate. We exploit this optimal solution to select the sets of slope and shift parameters for the proposed phase-frequency profile model (i.e. $\mathcal{M}$ and $\mathcal{I}$). Specifically, we propose an algorithm to select these sets that will be useful when deciding the unit cell design for a given RIS application. In this work, these sets are selected to increase the received signal strength over a wide range of directions. This is in contrast to existing works \cite{RISstableband, PracPhaseAmp, PracPhaseAmp2, PracPin}, where the set of possible RIS phase-frequency profiles are pre-determined by unit cell geometry.

    \item We then propose an algorithm to jointly optimize the power allocations across the sub-carriers and the phase shifts introduced by RIS unit cells based on the available phase-frequency profiles in order to maximize the rate. 
    
    %Unlike the previous work, where perfect channel estimation is assumed, we assume a known UE's location at the BS; this condition is realistic considering the localization algorithms and the good accuracy provided in many works \cite{RISLoc}.We compare our results against the benchmark scheme that assumes an ideal constant phase profile.
  
    \item We validate the performance of the proposed algorithms  through extensive numerical simulations. The results show an improvement in coverage under the proposed phase-frequency profiles compared to the scenario with idealized phase-frequency profiles.

     \item We then extend the analysis to the multi-UE MISO scenario. The algorithm to select the parameter sets $\mathcal{M}$ and $\mathcal{I}$ for the proposed RIS phase-frequency profile model in the SISO case is first extended. We then formulate and solve the optimization problem to design the sub-carrier power allocations and the unit cell phase-frequency profiles under zero-forcing and maximum ratio transmission precoding to maximize the sum rate. 
     
\end{itemize}

%that would improve the received signal strength over a range of angular directions
The rest of the paper is organized as follows. In Sec. \ref{rismodel}, we present a practical RIS unit cell design and a mathematical model for the associated phase-frequency profile.  Sec~\ref{systemModel} introduces the system model and problem formulation for the single-UE SISO OFDM scenario. Sec.~\ref{RISOptimizationandConfiguration} presents an algorithm to select the parameter sets for the proposed RIS phase-frequency profile model, and then proposes an algorithm to optimize the sub-carrier power allocation and the RIS phase-frequency profiles based on the designed sets to maximize the achievable rate. Sec.~\ref{resultSection} presents the numerical results for the SISO case. Sec~\ref{SecMISO} extends the analysis and results to the multi-UE MISO system, and Sec.~\ref{conclusion} concludes the paper.

\textit{Notations:} We denote matrices and vectors  by boldface upper-case and lower-case letters respectively, while $\mathbb{C}^{M\times N}$ is the space of $M \times N$ complex-valued matrices, and $\mathbb{Q}$ is the set of rational numbers. Also, $\mathrm{Card}(\mathcal{X})$, is the number of elements in set $\mathcal{X}$.  For a complex vector $\mathbf{a}$, $\lVert \mathbf{a} \rVert$, $\mathbf{a}^t$, and $\mathbf{a}^{h}$ denote the $l_2$-norm, transpose, and conjugate transpose respectively, while $\mathrm{diag}(\mathbf{a})$  denotes the diagonal matrix with main diagonal element equal to $\mathbf{a}$. Moreover $\mathbf{A}(i, :)$, $\mathbf{A}(:, j)$, and $\mathbf{A}(i, j)$ are the $i$-th row, the $j$-th column, and the $(i, j)$-th element of matrix $\mathbf{A}$ respectively, and $\mathbf{a}(i)$ denotes the $i$-th element of vector $\mathbf{a}$. The operations $\mathrm{mod}(.,.)$ and $\lfloor . \rfloor$ are the modulus and the floor arguments respectively.% Finally,  $j$ denotes the imaginary unit, i.e., $j^2 = -1$.

\section{Practical RIS Unit Cell Frequency Response} \label{rismodel}

In this section, we introduce an RIS unit cell design, and study the variation in the amplitude and phase  of the reflection coefficient with frequency for different unit cells dimensions. Based on the observations we make, we  propose a mathematical  model for the RIS phase-frequency profile that will be utilized in the subsequent sections.

\begin{figure}[!t]
\centering
\tikzset{every picture/.style={scale=.6}, every node/.style={scale=.9}}
\input{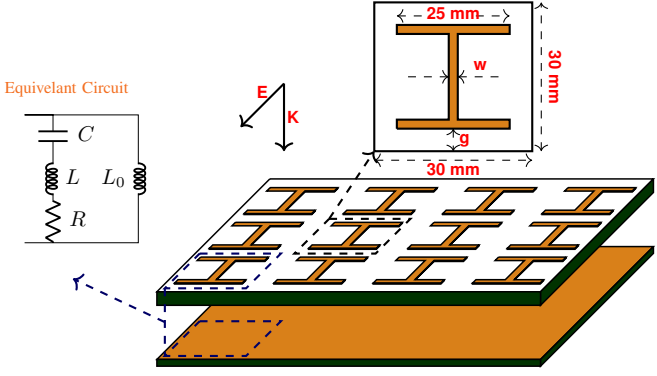}
\caption{An example of an RIS comprising of an array of resonant dog-bone elements placed over a ground plane. The unit cell geometry and the equivalent circuit are  shown in the inset diagrams. The orange areas are copper, the white areas are the isolating material, K is the direction of electromagnetic wave propagation, and E is the electrical field polarization.}
\label{RISCircuit}
\vspace{-0.5cm}
\end{figure}

The RIS consists of an array of conductive elements printed on an insulating material that provides a spatial filter response at a specific frequency by modifying the characteristics of the incoming electromagnetic waves. The shape of the conductive elements and the states of the active components, such as varactor diodes, embedded in the RIS surface determine the resonance frequency and $Q$ factor of the filter response of the RIS. Fig. \ref{RISCircuit} show an example of an RIS design with $N = 12$ unit cells. Each unit cell in the top layer has a dog-bone element placed over a ground plane.

%Since the RIS constitutes of a periodic repetition of  unit cells on the overall https://www.overleaf.com/project/6423245e3b6cb426b37c460csurface, the response of the surface can be characterized by studying the response of a single unit cell. 

\begin{figure}[!t]
   \centering
     \tikzset{every picture/.style={scale=.75}, every node/.style={scale=.9}}
    \input{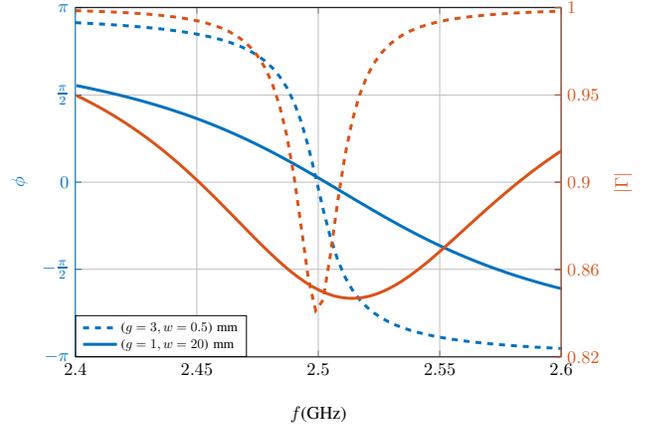}
   \caption{Amplitude- and phase-frequency responses of an RIS with unit cells of dimensions $g= 7$~mm and $w = 0.5$~mm (dashed curves) and $g= 0.6$~mm and $w = 25$~mm (solid curves).}
\label{ParallelUnitCell}
\vspace{-0.5cm}
\end{figure}

To characterize the response of the RIS, we investigate the reflection response of a single unit cell on the impinging electromagnetic waves in normal incidence as shown in Fig. \ref{RISCircuit}. An RIS element can be abstracted as an R-L-C series resonator, connected in parallel with an inductor of inductance $L_{0}$ (which is the equivalent inductance of the ground), with resistance $R$, inductance $L$, and capacitance $C$, that changes with the dimensions $g$ and $w$ of the element shown in Fig. \ref{RISCircuit}. 
%The amplitude and phase response of each element will therefore change with the shape of the element defined by these dimensions.  
We express the reflection coefficient of an RIS unit cell in Fig. \ref{RISCircuit} as  $\Gamma=|\Gamma|\exp(j\phi)$, and plot the amplitude $|\Gamma|$ and phase $\phi$ variation with frequency using the High-Frequency Simulation Software (HFSS). The results are shown in Fig. \ref{ParallelUnitCell}. Specifically, we plot the amplitude and phase-frequency profiles for the case where the dimensions of the unit cell are ($g = 7$, $w = 0.5$)~mm, which provides a fast-changing phase-frequency profile, and for the case where ($g = 0.6$, $w = 25$)~mm, which provides a slow-changing phase-frequency profile. Note that both cases have the same resonant frequency, with the former design having a larger inductance compared to the latter. Fig. \ref{ParallelUnitCell} clearly confirms that the reflection coefficients of the  RIS unit cells vary with frequency, which should be accounted for in the design of wide-band RIS-assisted systems. An example of an RIS surface with two pin diodes is presented in \cite{RISPin,RISRing}; the diodes are changing the effective length of the unit cell element resulting in phase reconfigurability.

Our work aims to provide a realistic simple model for the reflection coefficient of each RIS unit cell with a controllable phase-frequency profile and small reflection loss. It can be seen from Fig. \ref{ParallelUnitCell} that $|\Gamma| \in [0.85,1]$ (i.e. $|\Gamma| > -0.5$ dB) in the frequency band considered in the figure. To simplify the exposition in the rest of this paper, we consider the worst-case scenario and assume that $|\Gamma|$ = 0.85 over the whole band. 

Considering a constant amplitude-frequency response, we now develop a mathematical model for the controllable phase-frequency profile of each unit cell. 
%One can express the phase $\phi$ of the reflection coefficient $\Gamma$ as $\arctan\left(\frac{\Im(\Gamma)}{\Re(\Gamma)}\right)$. 
Based on observations from Fig. \ref{ParallelUnitCell}, we approximate the phase-frequency profile of $\Gamma$ as  \begin{equation}
    \label{eq15}
    \phi(f) = -2\arctan \small( m\times (f - f_{0}) + i_{0} \small),
\end{equation}
where $f_0$ is the center frequency of the OFDM band, $m$ controls the slope of the phase-frequency profile, and $i_0 = \tan \left(-\frac{\phi_{f_0}}{2}\right)$, where $\phi_{f_0}$ is the phase-shift at $f_0$.  

%In the proposed design, $|\gamma|$ of the RIS unit cells has a relatively small variation domain ($|\gamma| \in [0,0.7]$ dB).

%We assume a flat reflection coefficient with a reflection loss of 0.7 dB.It is similar to the proposed phase profile in \cite{RISAtan}, but with simpler parameters.

%The proposed $\arctan$ model is used to eliminate errors and misalignment between theoretical and practical RIS phase profiles.  Choosing the $\arctan$ function is  mathematically supported, as we consider the phase profile of a complex number $\Gamma$ in \eqref{eq101}, where $m$ is related to the geometrical shape of the element that is reflecting in the parallel unit cell, and $i_{0}$ is governed by the reverse voltage applied to the varactor diode in that element. Note that the element designs in the above figures show two extreme cases of phase response: the first (Fig. \ref{dogResponse}) with a large slope, i.e. large variation in the phase with frequency ($m = m_{max}$), and the second (Fig. \ref{patchResponse}) with a small slope ($m = m_{min}$). And then in each of these figures, phase reconfigurability at any given frequency is achieved by changing the diode capacitance that changes $i_0$. 

\begin{figure}[!t]
\centering
\tikzset{every picture/.style={scale=.8}, every node/.style={scale=.9}}
\input{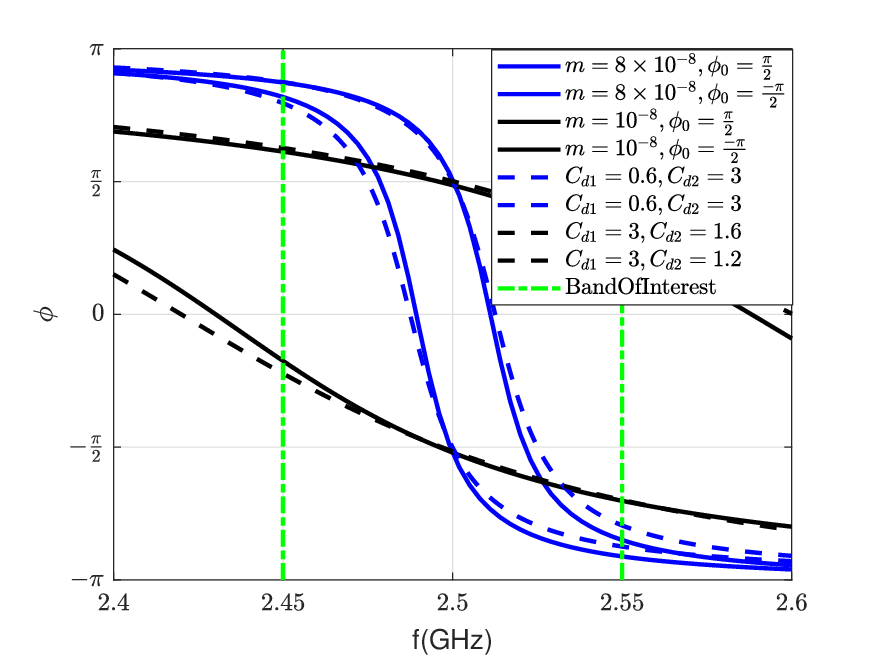}
\caption{Practical unit cell responses using HFSS and their approximations using \eqref{eq15}, for  two unit cell designs defined by dimensions $g$ and $w$ of ($6.8,0.5$)~mm and ($0.57,25$)~mm.}
\label{AtanApprox}
\vspace{-0.5cm}
\end{figure}

To illustrate the validity of the proposed phase-frequency profile model in \eqref{eq15}, we plot in Fig. \ref{AtanApprox} the HFSS simulated responses for two different RIS unit cell configurations shown in Fig. \ref{ParallelUnitCell}. We also plot their approximation using the $\arctan$ model in \eqref{eq15}, where the parameters $m$ and $i_0$ (stated in the figure) are selected to match the practical response. The proposed function in \eqref{eq15} is shown to provide a fairly tight and simple approximation to the practical phase-frequency response of the considered unit cell designs in the frequency band of interest (max phase error is 0.07 rad over the 100 \rm{MHz} band plotted in green on the figure). %It is worth mentioning that there are many works show the reconfigurability of the RIS by tuning the capacitance of an active element \cite{RISDesign_1,RISDesign_2}. 

In the rest of the paper, we make the following assumptions on the response of the RIS unit cells based on the design and model proposed in this section: \begin{itemize}
    \item The phase-frequency profile of each RIS unit cell can be modelled using \eqref{eq15}.
    \item The unit cell can provide a limited number of responses as determined by the possible values of slope $m$ and shift $i_0$. The set of slopes for each unit cell is represented as $\mathcal{M}$ and the set of shifts corresponding to slope $m$, with index $\tilde{m}$ in the set $\mathcal{M}$, is represented as $\mathcal{I}_{\tilde{m}}, \tilde{m} \in \{ 1,\dots,\mathrm{Card}(\mathcal{M}) \}$. In other words, each unit cell can select from a finite discrete set of phase-frequency profiles, independent of the profiles of the surrounding unit cells.
    \item Each unit cell can choose a single phase-frequency profile during a coherence block defined by $m$ and $i_{0}$.
\end{itemize}

Next, we present the system model and problem formulation under the practical phase-frequency profile model in \eqref{eq15} and the constraints defined above.

\vspace{-.1in}

\section{System Model} \label{systemModel}
This section presents the  system and channel models for  an RIS-assisted wide-band OFDM system, and formulates an achievable rate maximization problem under the proposed phase-frequency profile model in the last section. We focus on a single UE SISO system here, and extend the work to the multi-UE MISO system in Sec.~\ref{SecMISO}.

\vspace{-.05in}
\subsection{System description}

\begin{figure}[!t]
\centering
\tikzset{every picture/.style={scale=.65}, every node/.style={scale=.8}}
\newcommand{\wind}[3]{% args: label, coordinate                                                                                                                                                             
	\draw[black,line width=0.2mm] (#1) to (#2) to ($(#2)+(0,#3)$) to ($(#1)+(0,#3)$) to (#1);
}	

\newcommand{\beam}[4]{ %Color , capacity, angle
        \draw [fill=#1,opacity=#2] plot[variable=\t,domain=0:360,smooth,samples=51] 
        ({#3+18*sin(\t)}:{#4*pow(sin(\t/2),3)});
}

\newcommand{\CordinatesRIS}[0]{

        \draw[dashed,fill=white]  (12.4,9.05) circle (1.6);

        \draw[->,thick] (12.35,9) -- (13.05,9) node[below] {$y$};
        \draw[->,thick] (12.35,9) -- (11.85,8.65) node[below] {$x$};
        \draw[->,thick] (12.35,9) -- (12.35,9.8) node[left] {$z$};

         \draw[->,thick,yellow!50!black] (12.35,9) -- (12.85,9.5) node[right] {$UE$};

         \draw[dashed] (12.85,9.5) --  (12.8,8.7);
        \draw[dashed] (12.35,9) --  (12.8,8.7);
        
        \draw [->] ($(12.65,9.3)$) 
        arc (20:\ArcAngle:\ArcRadius);

        \node[black] at ($(12.75,9.75)$) {\scriptsize $\theta_{2}'$};

        \draw [->] ($(12.8,8.7)$) 
        arc (-45:\ArcAngleTwo:\ArcRadius);
        
        \node[black] at ($(12.35,8.4)$) {\scriptsize $\theta_{2}$};

         \node[black] at ($(11.55,9.05)$) {\scriptsize $x_{r},y_{r},z_{r}$};

}

\newcommand{\CordinatesBS}[0]{

        \draw[dashed,fill=white]  (12.7,9.05) circle (1.6);

        \draw[->, thick] (12.35,9) -- (13.05,9) node[below] {$y$};
        \draw[->, thick] (12.35,9) -- (11.85,8.65) node[below] {$x$};
        \draw[->, thick] (12.35,9) -- (12.35,9.8) node[left] {$z$};

         \draw[->, thick, red!60!black] (12.35,9) -- (12.85,9.5) node[right] {$RIS$};

         \draw[dashed] (12.85,9.5) --  (12.8,8.7);
        \draw[dashed] (12.35,9) --  (12.8,8.7);
        
        \draw [->] ($(12.65,9.3)$) 
        arc (20:\ArcAngle:\ArcRadius);

        \node[black] at ($(12.75,9.75)$) {\scriptsize $\theta_{1}'$};

        \draw [->] ($(12.8,8.7)$) 
        arc (-45:\ArcAngleTwo:\ArcRadius);
        
        \node[black] at ($(12.35,8.4)$) {\scriptsize $\theta_{1}$};

         \node[black] at ($(11.65,9.05)$) {\scriptsize $0,0,z_{b}$};

}

\newcommand*{\ArcAngle}{70}%
\newcommand*{\ArcAngleTwo}{-145}%
\newcommand*{\ArcRadius}{0.5}%

\begin{tikzpicture}
	%\node (b1) at (0,0) {};
	\node (b2) at (8,6) {};
	%\node (b3) at (5,-1) {};

     \node (bs) at (1.5,1.5) {};
    \node at ($(2.5,1)$) {\footnotesize Base-station};
    \draw[line width=1.5] ($(bs)$) to ($(bs)+(1,3)$) to ($(bs)+(2,0)$)to ($(bs)+(0.3,0.9)$)to ($(bs)+(1.5,1.5)$);
    \draw[line width=1.5]  ($(bs)+(0.48,1.44)$) to ($(bs)+(1.7,0.9)$) to ($(bs)$);
    \draw[line width=2] ($(bs)+(0.85,2.95)$) to ($(bs)+(1.15,2.95)$);
    \draw[line width=2] ($(bs)+(1,2.95)$) to ($(bs)+(1,3.08)$);

     \node (ris) at ($(b2)+(1.5,1.5)$) {};
    \node at ($(b2)+(1.5,3.2)$) {\footnotesize RIS};
    \draw[fill=blue!30!white] ($(b2)$) to ($(b2)+(3,0)$) to ($(b2)+(3,3)$) to ($(b2)+(0,3)$) to ($(b2)$);
        \draw ($(b2)+(0,0.6)$) to ($(b2)+(3,0.6)$);
        \draw ($(b2)+(0,1.2)$) to ($(b2)+(3,1.2)$);
        \draw ($(b2)+(0,1.8)$) to ($(b2)+(3,1.8)$);
        \draw ($(b2)+(0,2.4)$) to ($(b2)+(3,2.4)$);
        \draw ($(b2)+(0.6,0)$) to ($(b2)+(0.6,3)$);
        \draw ($(b2)+(1.2,0)$) to ($(b2)+(1.2,3)$);
        \draw ($(b2)+(1.8,0)$) to ($(b2)+(1.8,3)$);
        \draw ($(b2)+(2.4,0)$) to ($(b2)+(2.4,3)$);

	\node (u) at (13,3) {\ \ \footnotesize User};
	\draw[fill=gray!50!red] ($(u)+(0,.3)$) to ($(u)+(.25,.3)$) to ($(u)+(.3,.8)$) to ($(u)+(.05,.8)$) to ($(u)+(0,.3)$);

        \node[red] at ($(2,7)+(2.15,-0.2)$) {\scriptsize Transmitted};
	\node[red] at ($(2,7)+(2.15,-0.5)$) {\scriptsize beam};
	
	\node[yellow!30!red] at ($(ris)+(1.7,-2.5)$) {\scriptsize Redirected};
	\node[yellow!30!red] at ($(ris)+(1.7,-2.8)$) {\scriptsize beam};

         % axis units
        \draw[->,thick] (9.5,7.5) -- (10.25,7.5) node[below] {$y$};
        \draw[->,thick] (9.5,7.5) -- (9.5,8.25) node[right] {$z$};

         % Delta unit
        \draw[dashed] (8.3,9) -- (8.3,9.5) ;
        \draw[dashed] (8.9,9) -- (8.9,9.5) ;       
        \draw[<->] (8.3,9.15) -- (8.9,9.15) ;
        \node[black] at ($(8.6,9.3)$) {\scriptsize $\Delta$};

          \begin{scope}[shift={(2.5,4.5)}]
            \beam{red!50!white}{.7}{22}{5}
        \end{scope} 
    
         \begin{scope}[shift={(9.5,7.5)}]
            \beam{yellow!50!white}{.7}{-42}{3}
        \end{scope}

	\draw[->,thick,red!60!black] ($(bs)+(1,3)$) to (ris);
        \draw[->,thick,yellow!50!black] (ris) to ($(u)+             (.15,.9)$);

        %Magnify cordinates
        \draw[dashed]  (9.63,7.755) --  (12,10.4) ;
        \draw[dashed]  (9.63,7.48) --  (12,7.6) ;
        \draw [dashed]  (9.5,7.5) circle (0.25);
        \CordinatesRIS{};

        \draw [dashed] (2.5,4.5) circle (0.25);
        \draw[dashed]  (2.5,4.375) --  (5.8,1.2) ;
        \draw[dashed]  (2.5,4.625) --  (6.5,3.9) ;
         \begin{scope}[shift={(-6.5,-6.5)}]
            \CordinatesBS{}
        \end{scope}

\end{tikzpicture}
\caption{RIS-assisted system model.}
\label{SystemSetup}
\vspace{-0.5cm}
\end{figure}
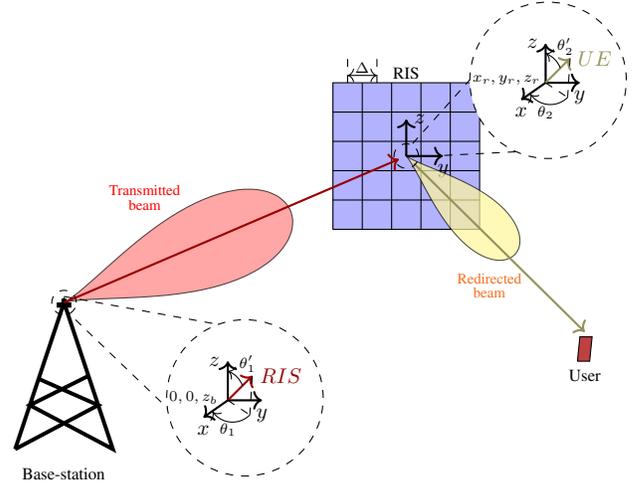

We study a downlink OFDM system, where a transmitter (Tx) serves a single UE through signals received from direct as well as reflected channels created by the RIS as shown in Fig. \ref{SystemSetup}. The Tx is fixed at coordinates $\small(0,0,z_{b}\small)$ meters, and  is equipped with one antenna with half-power beamwidths of $\theta_{\rm Tx}$ and $\phi_{\rm Tx}$ in the azimuth and elevation planes respectively. The UE is located at $\small(x_{u},y_{u},z_{u}\small)$ meters, and is equipped with a single  low-gain antenna with half-power beam widths of $\theta_{\rm UE}$ and $\phi_{\rm UE}$ in the azimuth and elevation planes respectively. %The antenna gains for the Tx and the UE are calculated using the formula provided in \cite{APHPBWNour}. 

An RIS is installed on a wall with its center at coordinates $ \small( x_{r},y_{r},z_{r} \small) $  meters to support the communication between the Tx and the UE while ensuring a line-of-sight (LoS) link between the Tx and the RIS. The RIS is a rectangular array with $N=N_y N_z$ unit cells, where $N_y$ is the number of unit cells along the y-axis and $N_z$ is the number of unit cells along the z-axis, arranged with a spacing of $\Delta$.  The position of element $j$ is $[x_{r},y_{r}+ \Delta n_{y,j},z_{r} + \Delta n_{z,j}]$, where $n_{z,j}$ and $n_{y,j}$ can be written for element $j$ as $\lfloor \frac{j - 1}{N_{y}} \rfloor + 1 -N_{z}/2$ and $\mathrm{mod}(j-1,N_{y}) + 1 - N_{y}/2$ respectively. The unit cells of the RIS redirect the signal toward the UE by introducing a controllable phase-shift value.  During a coherence block, each unit cell of the RIS  adopts one phase-frequency profile out of a given number of possible phase-frequency profiles (as discussed in Sec.~\ref{rismodel}), that can be selected using a microcontroller that gets this information from the  Tx via a control link.

%Note that we consider $2^{b/2}$ layers in a parallel unit cell resulting in $2^{b/2}$ different values of $m$ in \eqref{eq15}, and $2^{b/2}$  possible varactor diode capacitance values resulting in $2^{b/2}$ different values of $i_0$ in \eqref{eq15} for each value of $m$. Therefore the total number of responses for each RIS unit cell is $2^b$.

%Figure \ref{SystemSetup} shows an example of the described setup, where both the BS and the UE are on the side of the RIS; since we are studying an OFDM, 

\vspace{-.05in}
\subsection{Signal model} \label{signalmodel}

The OFDM signal consists of $N_{s}$ sub-carriers with a sub-carrier spacing $\Delta_f$ Hz. The channels are assumed to be flat within each sub-carrier band. The Tx utilizes the full bandwidth $N_{s} \Delta_{f}$ Hz to send data to the UEs. The transmitted signal $\mathbf{x}$ is given by $\mathbf{x} = [ x_{1} \sqrt{p_{1}},x_{2} \sqrt{p_{2}},\dots,x_{N} \sqrt{p_{N_{s}}}]$, where $x_{k}$ is the data symbol with unit power to be sent over sub-carrier $k \in {1,\dots,N_{s}}$, and  $p_{k}$ is the power allocated to sub-carrier $k$. The received base-band OFDM signal at the UE corresponding to sub-carrier $k$ can be written as
\begin{equation}
    y_{k} = \big (h_{k,d} +  g_{k} \big ) \sqrt{p_{k}}x_{k} + z_{k},  
   \label{eq7}
\end{equation}
where  $h_{k,d} \in \mathbb{C}$ is the Tx-UE direct channel, $g_{k} \in \mathbb{C}$ is the channel from the Tx to the UE through the RIS, and $z_{k} \in \mathbb{C}$ is the additive white Gaussian noise (AWGN) with zero mean and variance $\sigma^2$, i.e. $z_{k} \sim \mathcal{CN}(0,\sigma^2)$. The channel $g_{k}$ is given as $g_{k} = \mathbf{h}_{k,2}^{T} \boldsymbol{\Phi}_{k} \mathbf{h}_{k,1} $, where $\mathbf{h}_{k,1} = [h_{k,1,1},h_{k,1,2},\dots,h_{k,1,N}]^T \in \mathbb{C}^{N \times 1 }$ is the Tx-RIS channel, $\mathbf{h}_{k,2} = [h_{k,2,1},h_{k,2,2},\dots,h_{k,2,N}]^{T} \in \mathbb{C}^{ N \times 1}$ is the RIS-UE channel, and $\boldsymbol{\Phi}_{k} =  {\rm diag} \begin{pmatrix}
\gamma_{k,1} e^{j\phi_{k,1}}, \gamma_{k,2} e^{j\phi_{k,2}},\hdots, \gamma_{k,N} e^{j\phi_{k,N}} \end{pmatrix} \in \mathbb{C}^{N \times N} $ is a diagonal matrix representing the RIS response at sub-carrier $k$, where $\gamma_{k,i} = \gamma=0.85$ is the amplitude of the reflection coefficient as explained in Sec.~\ref{rismodel}, and $\phi_{k,i}$ is the phase shift introduced by unit cell $i$ at sub-carrier $k$ given by \eqref{eq15}.  Thus, we  write the RIS-assisted channel $g_{k}$ at sub-carrier $k$ as
\begin{equation}
\small
\begin{split}
    \label{eq11}
    g_{k} &= \gamma \sum_{j=1}^{N} h_{k,1,j} e^{j\phi_{k,j}} h_{k,2,j}
\end{split}
\end{equation}
%where $h_{k,1,j}$ and $h_{k,2,j}$ model the free-space path loss (FSPL) coefficients, the phase delays, and the fading for Tx-RIS and RIS-UE channels respectively, as outlined in the next subsection. Note that both $\alpha_{k,j}$ and $\beta_{k,j}$ are functions of the positions of the Tx, the RIS, and the UE, while the only controllable variable is $\phi_{k,j}$.
% \gamma   \begin{pmatrix}
% h_{k,2,1} 
%\hdots 
%h_{k,2,N} \\ 
%\end{pmatrix}
%\begin{pmatrix}
% e^{j\phi_{k,1}} & 0 & \hdots
% \\ \vdots & \ddots \\ 
% 0 & \hdots & e^{j\phi_{k,N}}
%\end{pmatrix} 
%\begin{pmatrix}
 %h_{k,1,1}
%\\\vdots \\
%h_{k,1,N} \\ 
%\end{pmatrix} \\
%    &= 

%The channel models are presented next.

 \vspace{-.05in}
 \subsection{Channel Model}\label{channelModel}

The direct channel $h_{k,d}$ at sub-carrier $k$  is modeled as
\begin{align}
    \label{eq12}
    h_{k,d}  &= \frac{c}{4\pi f_k d_{d} } G_{t} e^{-j \frac{ 2 \pi f_{k}}{c} d_{d}}  G_{r} + \chi_{k,d}, 
\end{align}
where $c$ is the speed of light, $f_{k} = f_{0} + \Delta_{f} (k - N_{s}/2)$ is the frequency of sub-carrier $k$, $d_{d}$ is the distance between the Tx and the UE,   $G_{t} (G_{r})$ is the Tx (UE) antenna gain, and $\chi_{k,d}$ represents the small-scale fading. We rewrite \eqref{eq12} as 
\begin{align}
    &h_{k,d} = \alpha_{k,d}e^{-j \nu_{k} d_{d}} + \chi_{k,d},
    \end{align}
    where  $\nu_{k} = \frac{ 2 \pi f_{k}}{c}$,  and $\alpha_{k,d}= \frac{G_{t} G_{r}}{2 \nu_{k}d_{d}}$  captures the free space path loss (FSPL) coefficient and antennas gain. 

Similarly, the channel between the Tx and element $j$ of the RIS at sub-carrier $k$ is modelled as
\begin{equation}
   \begin{split}
        \label{eq4}
    h_{k,1,j} = \frac{c}{4\pi f_{k} d_{1,j} } G_{t} e^{-j \nu_{k} d_{1,j}}  + \chi_{k,1,j},
   \end{split}
\end{equation}
where $d_{1,j}$ is the  distance between the Tx and  element $j$ of the RIS, and $\chi_{k,1,j}$ represents the small-scale fading in this link. Following the outdoor system assumption, we use the planar wave approximation to rewrite the channel coefficient as  \cite{PlannerWave}
\begin{align}
        \label{eq5}
    h_{k,1,j} & = \alpha_{k,1}  e^{-j \nu_{k} \Tilde{d}_{1,j}} + \chi_{k,1,j},
\end{align}
 where $\alpha_{k,1} = \frac{G_{t}}{2 \nu_{k}d_{1}}$ is the FSPL coefficient for the Tx-RIS channel multiplied by the Tx antenna gain, $d_{1} = \sqrt{(x_{b} - x_{r})^2 + (y_{b} - y_{r})^2 + (z_{b} - z_{r})^2}$ is the euclidean distance between the Tx and the center of the RIS, and  $ \Tilde{d}_{1,j} =  d_{1} + \Delta\small(n_{y,j} \cos{\theta_{1}}\sin{\theta_{1}'} + n_{z,j} \cos{\theta_{1}'} \small)$ is the approximated phase with respect to element $j$ at the RIS, with $\theta_{1}$ and $\theta_{1}'$ being the angles of arrival in the azimuth and elevation planes at the RIS respectively.  Similarly, the channel between RIS element $j$ and  UE over sub-carrier $k$ can be written as  %which can be written as
%\begin{equation}
 %       \label{eq6}
  %   \Tilde{d}_{1,i} =  d_{1} + \Delta\small(n_{y,i} %\cos{\theta_{1}}\sin{\theta_{1}'} 
   % + n_{z,i} \cos{\theta_{1}'} \small),
%\end{equation}
\begin{align}
        \label{eq27}
    h_{k,2,j} &= \alpha_{k,2}  e^{-j \nu_{k} \Tilde{d}_{2,j}} + \chi_{k,2,j},
\end{align}
where  $\alpha_{k,2} = \frac{G_{r}}{2 \nu_{k}d_{2}}$ is the FSPL coefficient for the RIS-UE channel multiplied by the UE antenna gain, $d_{2}$ is the euclidean distance between the center of the RIS and the UE, $ \Tilde{d}_{2,i} =  d_{2} + \Delta\small(n_{y,j} \cos{\theta_{2}}\sin{\theta_{2}'}+ n_{z,j} \cos{\theta_{2}'} \small)$, $\theta_{2}$ and $\theta_{2}'$ are the angles of departure in the azimuth and elevation planes respectively from RIS to the UE, and $\chi_{k,2,j}$ captures the small-scale fading.
    
We can further manipulate these expressions to write the RIS-assisted channels  $h_{k,1,j}$ and $h_{k,2,j}$ as
\begin{equation}\label{eq17}
    {h_{k,1,j}} =  \Tilde{\alpha}_{k,1,j} e^{-j \nu_{k} \big( d_{1} + \Delta \small( n_{y,j}\cos \theta_{1} \sin \theta_{1}' + n_{z,j}\cos \theta_{1}' \small)\big) + j \Tilde{\chi}_{k,1,j}}
\end{equation}
\begin{equation} \label{eq18}
    {h_{k,2,j}} = \Tilde{\alpha}_{k,2,j} e^{-j \nu_{k} \big(d_{2} + \Delta \small( n_{y,j}\cos \theta_{2} \sin \theta'_{2}  + n_{z,j}\cos \theta'_{2} \small)\big) + j\Tilde{\chi}_{k,2,j}}
\end{equation}
where $\Tilde{\chi}_{k,1,j}$ ($\Tilde{\chi}_{k,2,j}$) and $\Tilde{\alpha}_{k,1,j}$ ($\Tilde{\alpha}_{k,2,j}$) represent the phase and amplitude of $\alpha_{k,1} + \chi_{k,1,j} e^{j\nu_{k} \Tilde{d}_{1,j} }$ ($ \alpha_{k,2} + \chi_{k,2,j} e^{j\nu_{k} \Tilde{d}_{2,j} }$), respectively. The received signal in \eqref{eq7} at sub-carrier $k$ can now be written  as \vspace{-.1in}
\begin{equation}\label{eq16}
    y_{k} =  r_{k} \sqrt{p_{k}}x_{k} + z_{k}, 
\end{equation}
where
\begin{align}\label{eq19}
    r_{k} &=\alpha_{k,d} e^{-jd_{d} \nu_{k}} + \chi_{k,d} + e^{- j d \nu_{k}} \nonumber \\
        & \sum_{j=1}^{N}  \tau_j e^{-j \nu_{k} \small( \Delta_{y,j} + \Delta_{z,j}\small) +j \small( \Tilde{\chi}_{k,j} + \phi_{k,j} \small)},
\end{align} 
%\begin{align}
%    &r_{k} = h_{k,d} + g_{k}=h_{k,d}+ \gamma \sum_{i=1}^{N} \alpha_{k,i}\phi_{k,i}\beta_{k,i},
%\end{align}
%is the aggregate channel between the Tx and the UE, which can be written using the models above  as
where $\tau_j =  \Tilde{\alpha}_{k,1,j} \Tilde{\alpha}_{k,2,j}  \gamma$,  $ \Tilde{\chi}_{k,j} = \Tilde{\chi}_{k,1,j}+\Tilde{\chi}_{k,2,j}$,  $d = d_{1} + d_{2}$, $\Delta_{y,j} = n_{y,j} \Delta \small( \cos \theta_{1} \sin \theta_{1}' + \cos \theta_{2} \sin \theta_{2}'   \small)$, and $\Delta_{z,j} = n_{z,j} \Delta \small(  \cos\theta_{1}' + \cos\theta_{2}' \small)$. Note that the term $\Tilde{\chi}_{k,j}$ in the overall channel model is random as it captures the effect of small-scale fading. In the next section, when selecting the sets for the parameters $m$ and $i_0$ describing the RIS phase-frequency profile in \eqref{eq15}, and then optimizing the RIS phase shifts using the designed sets, we will assume $\Tilde{\chi}_{k,i}=0$. The reason for doing this when designing the RIS parameter sets is to select these sets based on the geometrical information related to Tx, RIS and UEs, in order to have a generic framework that informs the best RIS design that can be used later for RIS fabrication and deployment. The reason for assuming $\Tilde{\chi}_{k,i}=0$ in the RIS phase-shift optimization stage (after deployment) is to make the optimization location dependent only, thereby reducing the need for full channel state information acquisition, and because it is not feasible to reconfigure the RIS phase-shifts at the pace of small-scale fading.  However, we will integrate the effect of small-scale fading  $\Tilde{\chi}_{k,i}$ in the numerical results later, and show that the proposed design performs well under fading as well.

%The expressions in \eqref{eq19} will be used to find the optimum response of the RIS elements to maximize the achievable rate. 

\vspace{-.3in}
\subsection{Problem Formulation} \label{Cpacityformula}

The achievable rate at the user is given as 
\begin{equation}
    \label{eq13}
    R = \sum_{k=1}^{N_{s}} \log_{2} \left( 1+\frac{ 
 | r_{k} |^2 p_{k}}{\sigma^{2}\Delta_f} \right).
\end{equation} 
where $\sigma^{2}$ is the noise power. The rate in \eqref{eq13} is a function of the RIS phase-frequency profiles $\phi_{k,j}$, $j=1,\dots, N$, $k=1,\dots, N_s$ that are given by \eqref{eq15} and are parameterized by $m$ and $i_{0}$. Each  RIS element can choose a phase-frequency profile from a discrete set of available profiles specified by sets $ \mathcal{M}$ and $\mathcal{I}_{\tilde{m}}$ of slopes $m$ and shifts $i_0$ respectively as discussed towards the end of Sec. II. Specifically,  element $j$ can choose a value of slope $m_j$ from the set $ \mathcal{M}$ where $\mathrm{Card}(\mathcal{M}) = 2^{b/2}$, and a value of shift $i_{0j}$ corresponding to slope $m_j$ from the set $ \mathcal{I}_{\tilde{m}}$ where  $\mathrm{Card} (\mathcal{I}_{\tilde{m}}) = 2^{b/2}$ and $\tilde{m}$ is the index of selected slope $m_j$ in set $\mathcal{M}$, i.e. $\tilde{m} \in \{1 ,\dots,2^{b/2}\}$. Therefore the total number of available responses at each RIS element is $2^b$. Defining $\mathbf{p}=[p_1, \dots, p_{N_s}]$ as the power allocation vector and $\boldsymbol{\Phi}= \text{diag}(\boldsymbol{\Phi}_{1}, \boldsymbol{\Phi}_{2}, \dots,\boldsymbol{\Phi}_{N_s}) \in \mathbb{C}^{NN_s\times NN_s}$ as the RIS phase shifts matrix where $\boldsymbol{\Phi}_k=\text{diag}(\phi_{k,1},\dots, \phi_{k,N})\in \mathbb{C}^{N\times N}$ is the phase-shifts matrix at sub-carrier $k$, we   formulate the following optimization problem:
\begin{subequations}
\label{eq14}
 \begin{alignat}{2} \hspace{-.09in} \textit{(P1)} \hspace{.1in}
&\!\max_{\begin{subarray}{c}
  \boldsymbol{\Phi},
  \bold{p}
  \end{subarray}} \hspace{.1in}         && \frac{1}{N_{s}} \sum_{k=1}^{N_{s}} \log_{2} \left(1+\frac{ 
 |r_{k}|^2 p_{k}}{\sigma^{2} \Delta_f} \right) \label{obj1}\\
&\text{subject to} \hspace{.05in} &      & \sum_{k=1}^{N_{s}} p_{k} \leq P, \\
% && &   ||\gamma_{n,k} ||  = \gamma, \\
&&      & \phi_{k,j} =-2\text{$\arctan$}(m_{j}(f_k -f_{0})+i_{0j})   \nonumber \\
&& & \text{for } j=1,\dots, N, k=1,\dots, N_s \label{ann} \\
& & & m_{j}\in \mathcal{M} \label{constrainti2} \\
& & &  i_{0j}\in \mathcal{I}_{\tilde{m}}, \tilde{m}=\text{index}(m_j) \in \{1 ,\dots,2^{b/2}\} \label{constrainti}
\end{alignat}
\end{subequations}
%where  $\mathcal{M}$ and $\mathcal{I}$ are the sets of $2^{b/2}$ possible values of $m$ and $i{0}$ respectively that a parallel unit cell can choose from by deciding the element that is reflecting and the reverse bias voltage applied to the varactor diode in that element.
% Subjected to $\sum_{k=1}^{N_{s}} p_{k} \leq P, \; \lVert \gamma_{k,i} \lVert  = \gamma, \; \forall{k} \in [1,N_{s}]$, $\forall i \in [1,N]$. Although the water-filling algorithm has some advantages, it will not be utilized while we optimize the RIS response. 

For given  $\phi_{n,k}$'s, the power allocation can be found optimally using water-filling. Optimizing the response of the RIS is more complicated, as OFDM channel responses as well as the phase-frequency relationship in \eqref{eq15}, need to be accounted for. Moreover, before solving \textit{(P1)}, we need to select the sets $\mathcal{M}$ and $\mathcal{I}_{\tilde{m}}$ for the RIS unit cells that would yield a favourable design for rate maximization. The method employed for the selection of sets $\mathcal{M}$ and $\mathcal{I}_{\tilde{m}}$, and for solving Problem \textit{(P1)} are discussed in detail in the next section. It is worth mentioning that the selection of the sets $\mathcal{M}$ and $\mathcal{I}_{\tilde{m}}$ and the solution of the optimization problem \textit{(P1)} will be based on knowledge of only  the geometrical locations of the Tx and RIS, and the set of possible locations of the UE. 

%A comparison between these two \textcolor{red}{ the equal power allocation and the water-filling algorithms} algorithms can be found in the results (Section \ref{resultSection}). 

\vspace{-.05in}
\section{Proposed RIS Design and Optimization}
\label{RISOptimizationandConfiguration}

 As discussed earlier in Sec.~\ref{rismodel}, the phase-frequency profile of each unit cell is represented by the proposed $\arctan$ model in \eqref{eq15} that is parametrized by   $m$ and $i_0$. Further,  these parameters can take values from discrete sets $\mathcal{M}$ and $\mathcal{I}_{\tilde{m}}$ of size $2^{b/2}$ each. In this section, for a given scenario (known positions and physical properties of Tx and RIS), we first propose a method to select the sets $\mathcal{M}$ and $\mathcal{I}_{\tilde{m}}$ of the possible values of $m$ and $i_0$ respectively, resulting in an optimized RIS design with unit cells that can provide $2^b$ possible phase-frequency profiles. Later, we propose an algorithm to solve \textit{(P1)} based on the selected  sets $\mathcal{M}$ and $\mathcal{I}_{\tilde{m}}$. 

%The first step in the design of our RIS-assisted OFDM system under the phase frequency relationship in \eqref{eq15}, is to find  optimal values of RIS parameters $m$ and $i_0$ for the considered channel model in \eqref{eq19} for any given UE location and use this to build the sets $\mathcal{M}$ and $\mathcal{Q}$. Then based on these sets, an algorithm to solve \textit{(P1)} will be proposed. 

%which is characterized by $m$ and $i_{0}$, which can take a value from predefined values limited by the number of layers (the number of quantization levels of the reverse voltage).

%The $\arctan$ function used to model the RIS response saturates at $\pi$ and $-\pi$, making the optimization problem more complex.

%In this work, we model the RIS response with an $\arctan$ function with varying transition speeds ($m$) and zero-crossing phases ($i_0$). The values of these parameters will be chosen based on the channel and RIS structure. The following sections addresses the selection of RIS parameters and outlines the optimization problem and proposed algorithms for configuring and optimizing RIS.

\vspace{-.1in}
\subsection{RIS Design: Selection of Sets $\mathcal{M}$ and $\mathcal{I}_{\tilde{m}}$} \label{RISPara}

%In this section, we build the RIS design dictated by the sets $\mathcal{M}$ and $\mathcal{I}$ that would be favourable for achievable rate maximization at a UE in a SISO-OFDM system under the channel model in \eqref{eq19}. The extension of the proposed design to a multi-UE MISO case appears in the next section. Once the sets have been built and the RIS is fabricated, the Tx can only choose parameters from these sets to control the frequency-dependent phase profiles that each parallel unit cell of the RIS can induce. This is in contrast to 

Most works on RIS-assisted systems assume a given RIS structure, i.e. a  given set of discrete phase shifts that the RIS unit cells can apply, and study the system under this assumption. In practice, we can design an RIS with unit cells that provide a desired set of phase-frequency profiles. In this subsection, we devise a method to select the set of possible phase-frequency profiles, i.e. the sets $\mathcal{M}$ and $\mathcal{I}_{\tilde{m}}$, that the RIS should be designed to provide in order to maximize the gain from using an RIS in terms of the received signal strength over a desired geographical region. We remark here that our proposed method for selecting the sets $\mathcal{M}$ and $\mathcal{I}_{\tilde{m}}$ will only utilize the geometrical locations of the Tx and RIS, and the set of possible locations of UE which is expected to be defined for an RIS-assisted system (where an RIS is deployed to serve a specific area). Restricting the attention to the geometric characteristics of the channel serves the goal of making the resulting RIS design generic and applicable in scenarios which share the same geometry. 

To find the sets $\mathcal{M}$ and $\mathcal{I}_{\tilde{m}}$, we first derive the optimal phase-frequency profile based on the channel model in  \eqref{eq19} that maximizes the received signal strength (and, in turn, the achievable rate) at a given UE location, and then use the resulting expression to outline a method for developing the sets $\mathcal{M}$ and $\mathcal{I}_{\tilde{m}}$. To achieve the maximum received signal power, it is necessary that the direct and reflected signals add constructively at the  UE.  This condition can be expressed for unit cell $j$ using \eqref{eq19} as
\begin{equation}\label{eq20}
  \begin{aligned}
        d_{d} \nu_{k} = -\nu_{k} \big( d + \Delta_{y,j} + \Delta_{z,j} \big) +  \phi_{k,j}
  \end{aligned}
\end{equation}

%Following the derivation in Sec.~\ref{channelModel}, the parameters affecting this relationship are the Tx-RIS and RIS-UE distances, the  angles describing the Tx-RIS and RIS-UE links, the center frequency $f_{0}$,  the frequency of each sub-carrier, and the inter-element separation. 

%containing possible values of $m$ and $i0$ that would control the phase-shift given by the $\arctan$ function in \eqref{eq15} that each unit cell can apply.

%\subsubsection{LoS and ILoS received signal}\label{ILOSLOSRIS} the effecting variables like the distance of the BS-RIS and RIS-UE links,  angle of arrival to the RIS, angle of departure from the RIS, and the number of OFDM sub-carriers.

%Following the derivation in section \ref{channelModel}, the parameters affecting the received signal phase are the distance of the BS-RIS and RIS-UE links, and  angles describing the BS-RIS and RIS-UE links. To analyze the behaviour of the received signal phase, we perform a characterization of the parameters' influence on the received signal phase. Using this analysis, we define the parameters of the functions in equation \eqref{eq15}. 

%In order to do this, we can write an expression for the reflected phases through the elements of the RIS based on the condition for constructive summation.

Let the spacing between unit cells be given by $\Delta = \zeta \lambda_{0}$, where $\lambda_{0}$ is the wavelength corresponding to $f_{0}$ and $\zeta \in [0.5,1]$ in order to minimize the coupling between unit cells.  Using \eqref{eq20}, the optimal phase shift $\phi_{k,j}$ that should be introduced by unit cell $j$ at sub-carrier $k$ can be written as 
\begin{align}\label{eq21}
        \phi_{k,j}^* =   \Big(& 2\pi \zeta \frac{f_{k}}{f_{0}}  \big(\Delta_{y,j}+ \Delta_{z,j} )\nonumber \\
        &  +\nu_{k}(d_{d} - d_{1} -d_{2})\Big)\mathrm{mod}(2\pi)
\end{align}

The solution in \eqref{eq21} is a linear function of frequency $f_{k}$ (that repeats every $2\pi$ radians) with different slopes and intercepts for different element index $j$ as well as for different  RIS and UE locations. Note that we can account for the effect of $(d_{d} - d_{1} -d_{2}) \nu_{k}$ in \eqref{eq21}  using precoding at the Tx since this factor is common for all elements. Consequently, we can write \eqref{eq21} compactly as \vspace{-.05in}
\begin{equation}
    \label{eq25}
    \phi_{k,j}^* = -{a}^*_{j} (f_{k}-f_{0}) + b^*_{0j},
\end{equation}
where ${a}^*_{j}$ and ${b}^*_{0j}$ are the  slope and intercept of the optimal linear phase-frequency relationship in \eqref{eq21}, for a certain position of Tx, UE and RIS element $j$ and are given as 
\begin{align}\label{eq26}
        &{a}^*_{j} =  \frac{2\pi \zeta}{f_{0}}  \big(\Delta_{y,j} + \Delta_{z,j} \big) , \\ 
\label{eq107}
        &{b}^*_{0j}= 2\pi \zeta  \big(\Delta_{y,j} + \Delta_{z,j} \big) \mathrm{mod}(2\pi)   
\end{align}

Note that while the slope of the phase response function with respect to frequency in \eqref{eq21} is positive, we used $-{a}^*_{j}$ instead of ${a}^*_{j}$ in \eqref{eq25}. This change in sign for all RIS elements will not change the received signal power $|r_k|^2$ in \eqref{obj1} as the relative phase difference between elements remains the same. The reason we change the sign is to allow us to match the direction of the slope of the optimal phase-frequency profile in \eqref{eq21}  with that of the phase-frequency profile of a practical RIS design shown in Fig. \ref{ParallelUnitCell} that has a negative slope. 

%The values of the slope is a function of center frequency,  position of element, and the angles of arrival and departure, while the intercepts are functions of the same parameters in addition to $d_{d}$ and $\Tilde{d}$, which make them random variables.  
There are two main issues that make the solution in \eqref{eq25} impractical to implement: firstly, the sets of values of ${a}^*_j$ and $b^*_{0j}$ corresponding to all possible UE locations, i.e. $\frac{\pi}{2} \leq \theta_2 \leq \frac{3\pi}{2}$ and $0\leq  \theta_2'\leq \pi$, are continuous, while we can only have a discrete set of possible phase-frequency profiles at each unit cell specified by the sets $\mathcal{M}$ and $\mathcal{I}_{\tilde{m}}$.  Therefore we need to map the infinitely many values of ${a}^*_j$'s and $b^*_{0j}$'s corresponding to all UE directions to values in discrete sets. Secondly, the practical RIS's phase-frequency profile saturates at  $\pi$ and $-\pi$, as shown in Fig. \ref{ParallelUnitCell}. On the other hand,  the optimal response in \eqref{eq25} wraps around these values. 

\begin{figure}[!t]
\centering
\tikzset{every picture/.style={scale=.8}, every node/.style={scale=.9}}
\begin{tikzpicture}

\begin{axis}[%
width=3.75in,
height=3in,
at={(0.758in,0.497in)},
scale only axis,
xmin=2.4,
xmax=2.6,
xtick={2.4,2.45,2.5,2.55,2.6},
xticklabels={{$2.4$},{$2.45$},{$2.5$},{$2.55$},{$2.6$}},
xlabel style={at={(axis cs:2.5,-3.25)},anchor=north},
xlabel={$f_{k}$(GHz)},
ymin=-3.14259265358979,
ymax=3.14259265358979,
ytick={-3.14159265358979,-1.5707963267949,0,1.5707963267949,3.14159265358979},
yticklabels={{$-\pi$},{$-\frac{\pi }{2}$},{$0$},{$\frac{\pi}{2}$},{$\pi$}},
ylabel={$\phi(f_{k})$},
ylabel style={at={(axis cs:2.397,0)},anchor=north},
axis background/.style={fill=white},
xmajorgrids,
ymajorgrids,
legend style={{at =  (axis cs:2.4,-3.1415)}, nodes={scale=0.75, transform shape}, anchor=south west, legend cell align=left, align=left, draw=black}
]
\addplot [color=black, dashed, line width=1.5pt]
    table[row sep=crcr]{%
2.35	0\\
2.36	-0.418879020478639\\
2.37	-0.837758040957278\\
2.38	-1.25663706143592\\
2.39	-1.67551608191456\\
2.4	-2.0943951023932\\
2.41	-2.51327412287183\\
2.42	-2.93215314335047\\
2.43	2.93215314335047\\
2.44	2.51327412287183\\
2.45	2.0943951023932\\
2.46	1.67551608191456\\
2.47	1.25663706143592\\
2.48	0.837758040957278\\
2.49	0.418879020478639\\
2.5	0\\
2.51	-0.418879020478639\\
2.52	-0.837758040957278\\
2.53	-1.25663706143592\\
2.54	-1.67551608191456\\
2.55	-2.0943951023932\\
2.56	-2.51327412287183\\
2.57	-2.93215314335047\\
2.58	2.93215314335047\\
2.59	2.51327412287183\\
2.6	2.0943951023932\\
2.61	1.67551608191456\\
2.62	1.25663706143592\\
2.63	0.837758040957278\\
2.64	0.418879020478639\\
2.65	0\\
};
\addlegendentry{$ \phi^{*} = 24 \times 10^{-6}   (f - f_{0})$}

\addplot [color=black, line width=1.5pt]
 table[row sep=crcr]{%
2.35	2.62038787009511\\
2.36	2.58499333557957\\
2.37	2.54459479041744\\
2.38	2.49809154479651\\
2.39	2.44405064642198\\
2.4	2.38057989936506\\
2.41	2.30514399443134\\
2.42	2.21429743558818\\
2.43	2.10330042509675\\
2.44	1.96558744649466\\
2.45	1.79211076914269\\
2.46	1.5707963267949\\
2.47	1.28700221758657\\
2.48	0.927295218001612\\
2.49	0.489957326253728\\
2.5	-0\\
2.51	-0.489957326253728\\
2.52	-0.927295218001612\\
2.53	-1.28700221758657\\
2.54	-1.5707963267949\\
2.55	-1.79211076914269\\
2.56	-1.96558744649466\\
2.57	-2.10330042509675\\
2.58	-2.21429743558818\\
2.59	-2.30514399443134\\
2.6	-2.38057989936506\\
2.61	-2.44405064642198\\
2.62	-2.49809154479651\\
2.63	-2.54459479041744\\
2.64	-2.58499333557957\\
2.65	-2.62038787009511\\
};
\addlegendentry{$\phi = -2 \arctan \left( 2.5\times 10^{-8} (f - f_{0})\right)$}

\addplot [color=blue, dashed, line width=1.5pt]
  table[row sep=crcr]{%
2.35	1.59697626557481\\
2.36	1.59523093632282\\
2.37	1.59348560707082\\
2.38	1.59174027781883\\
2.39	1.58999494856683\\
2.4	1.58824961931484\\
2.41	1.58650429006285\\
2.42	1.58475896081085\\
2.43	1.58301363155886\\
2.44	1.58126830230686\\
2.45	1.57952297305487\\
2.46	1.57777764380287\\
2.47	1.57603231455088\\
2.48	1.57428698529889\\
2.49	1.57254165604689\\
2.5	1.5707963267949\\
2.51	1.5690509975429\\
2.52	1.56730566829091\\
2.53	1.56556033903891\\
2.54	1.56381500978692\\
2.55	1.56206968053493\\
2.56	1.56032435128293\\
2.57	1.55857902203094\\
2.58	1.55683369277894\\
2.59	1.55508836352695\\
2.6	1.55334303427495\\
2.61	1.55159770502296\\
2.62	1.54985237577096\\
2.63	1.54810704651897\\
2.64	1.54636171726698\\
2.65	1.54461638801498\\
};
\addlegendentry{$\phi^{*} = 10^{-8} \times  (f - f_{0}) + \frac{\pi}{2}$}

\addplot [color=blue, line width=1.5pt]
 table[row sep=crcr]{%
2.35	1.71344442777789\\
2.36	1.70482346203696\\
2.37	1.69611663540381\\
2.38	1.68732299375829\\
2.39	1.67844157728648\\
2.4	1.66947142076658\\
2.41	1.66041155387441\\
2.42	1.651261001509\\
2.43	1.64201878413927\\
2.44	1.63268391817257\\
2.45	1.62325541634585\\
2.46	1.61373228814028\\
2.47	1.60411354022032\\
2.48	1.59439817689791\\
2.49	1.5845852006228\\
2.5	1.57467361249984\\
2.51	1.56466241283407\\
2.52	1.55455060170467\\
2.53	1.54433717956843\\
2.54	1.53402114789391\\
2.55	1.52360150982691\\
2.56	1.51307727088839\\
2.57	1.50244743970557\\
2.58	1.49171102877712\\
2.59	1.4808670552733\\
2.6	1.46991454187199\\
2.61	1.45885251763125\\
2.62	1.44768001889932\\
2.63	1.43639609026284\\
2.64	1.42499978553394\\
2.65	1.4134901687769\\
};
\addlegendentry{$\phi = -2 \arctan \left(  10^{-9} \times  (f - f_{0}) + 1\right)$}

\node[draw,scale = 0.75,align=left,anchor=north east] at (axis cs: 2.57,3.14 ) {Solid:  $\arctan$ profile \eqref{eq15} \\Dashed: optimal linear profile \eqref{eq25}};
\end{axis}
\end{tikzpicture}%
\caption{Optimal phase-frequency response $\phi^*$ in \eqref{eq25} and the nearest practical $\arctan$ phase-frequency response  $\phi$ in \eqref{eq15}.}
\label{SubSetResponse}
\vspace{-0.5cm}
\end{figure}
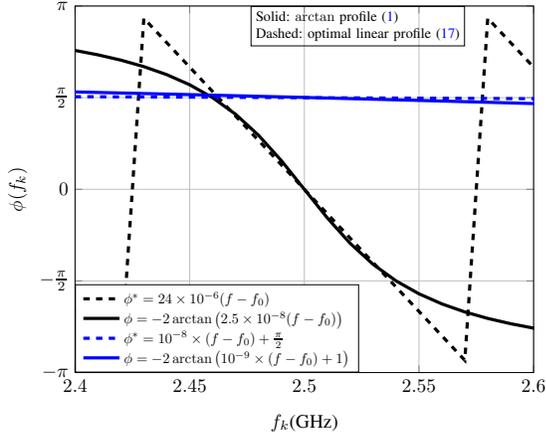

To highlight this further, we plot $\phi_{k,j}^*$ in \eqref{eq25} against frequency in Fig. \ref{SubSetResponse} for the first and last elements of a linear RIS considering $\theta_1= \pi$, $\theta_1'= \frac{\pi}{2}$, $\theta_2=\frac{3\pi}{4}$, $\theta_2'=\frac{\pi}{2}$, and $N = 200$. The optimal values of $({a}^*_j,b^*_{0j})$ for both elements are calculated and mentioned on the figure as $(24\times 10^{-6},0 )$ and $(10^{-8},\pi/2 )$. We also plot the closest $\arctan$ phase-frequency profile approximation in \eqref{eq15} for each of the two cases. We do so by choosing the best value of the pair $(m_j,i_{0j})$ that parameterizes the phase-frequency model in \eqref{eq15} such that $ m_j\in \{m_{\rm min}, m_{\rm max}\}$ and $i_{0j}\in \{-1, 1\}$. The pairs are selected numerically to minimize the mean squared error (MSE) between the optimal linear phase-frequency profile in \eqref{eq25} and the $\arctan$ phase-frequency profile in \eqref{eq15} over the simulated frequency band. The values of $(m_j, i_{0j})$ turn out to be $(2.5\times 10^{-8},0 )$ and $(10^{-9},1)$ for the two considered elements. Note that the values of $(m_j, i_{0j})$ for the two considered elements under the practical $\arctan$ phase-frequency profile model in \eqref{eq15} that yield a good match to the optimal phase response in \eqref{eq25} are different from ${a}_j^*$ and $b^*_{0j}$. We can also observe from this figure that  the optimal linear phase profile does not saturate at $\pi$ and $-\pi$ but rather wraps around these values, i.e. the response repeats periodically,  while the practical phase-frequency profile saturates at $\pi$ and $-\pi$.

%Further in figure \ref{CMFSlope}, we plot the CDF of the optimal slope shows an example of optimum slopes Probability Mass Function (PMF) for different $\theta_{1}$ values, where changing $\theta_{1}$ results in different distribution and maximum slope value. The above relation shows the importance of choosing a proper value of $\mathbf{a}$ in \eqref{eq15} of RIS elements. 

%Therefore we need to carefully design the  

%\begin{figure}[!t]
%\centering
%\includegraphics[width=\linewidth]{Figures/SystemModel/slope-intercept.eps}
%\caption{Optimum slope-intercept pairs for different $\theta_{2}$ values while keeping $\theta_{1}$, $\theta_1'$, and $\theta_2'$ fixed.}
%\label{SlopeIntercept}
%\end{figure}

%We now build the RIS parameters of the practical phase response model  in \eqref{eq15}outlined in Sec.~\ref{rismodel} under constraints of a limited number of total responses. The optimal solution for  problem \textit{(P1)} is to build each RIS element with a different slope and intercept  that changes with channel conditions according to \eqref{eq21}. To highlight this, we write this solution in \eqref{eq21} compactly for each RIS element as

Thus, we need to select a set of $\arctan$ phase-frequency profiles given by \eqref{eq15} to match as closely as possible the optimal theoretical solution. However, the optimal solution depends on the UE location, and we need the RIS to be useful for different UE locations. To approach a solution that achieves this objective, we calculate all  slope-intercept pairs corresponding to the optimal linear phase response in \eqref{eq25} for all $N$ elements of the RIS, while fixing the Tx and the RIS locations and changing the UE's location, i.e. choosing $\theta_{2} \in \{\frac{\pi}{2},\frac{3 \pi}{2}\}$ and $\theta_{2}' = \{ 0,\pi \}$ with a step size of $\Delta_{\theta}$.  The set of slope-intercept pairs $\mathcal{S}$ for given $\theta_{1}$ and $\theta_{1}'$ can be written as follows
\begin{equation}
\label{eq140}
    \mathcal{S}_{\theta_{1},\theta_{1}'} =  \left\{ \left( a^{*}_{j},b^{*}_{0j} \right) \left|
    \begin{subarray}{c}
        j \in \{1,\dots,N \} \\
        \theta_{2} = \{ \pi/2:\Delta_{\theta}:3\pi/2 \} \\
        \theta_{2}' = \{ 0:\Delta_{\theta}:\pi \}
    \end{subarray}
    \right. 
  \right\}
\end{equation}

Note that changing $\theta_{1}$ or $\theta_{1}'$ will result in a (qualitatively) similar set. The next step is to select the set of  slopes $\mathcal{A}$  with $\mathrm{Card}(\mathcal{A})=2^{b/2}$, as well as the set of  intercepts $\mathcal{B}_{\tilde{a}}$ with  $\mathrm{Card}(\mathcal{B}_{\tilde{a}})=2^{b/2}$ corresponding to  slope index $\tilde{a}$, where $ \tilde{a} \in \{1 ,\dots,2^{b/2}\}$. These discrete sets will contain the parameters  $a$ and  $b_{0}$ chosen from the optimal continuous set of parameters $(a^{*}_{j},b^{*}_{0j})$ that were obtained corresponding to all RIS elements, i.e. $j \in \{1,\dots,N\}$  and all user locations as defined in \eqref{eq140}. Then using the sets $\mathcal{A}$ and $\mathcal{B}_{\tilde{a}}, \tilde{a} \in \{1 ,\dots,2^{b/2}\}$, we find the $\arctan$ phase-frequency profile parameter sets $\mathcal{M}$ and $\mathcal{I}_{\tilde{m}}, \tilde{m} \in \{1 ,\dots,2^{b/2}\}$.

%The optimum RIS configuration problem is two independent problems for slope and intercept. 

\begin{figure}[!t]
   \begin{minipage}{0.48\textwidth}
     \centering
     \tikzset{every picture/.style={scale=.8}, every node/.style={scale=.9}}
    \input{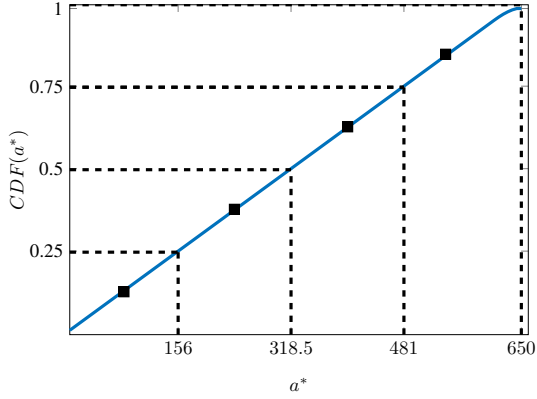}
    \subcaption{CDF of the optimal slope $a_j^*$  in \eqref{eq26} for all elements and users' locations. The abscissa of the black squares represent the values of $a$  chosen to constitute the set $\mathcal{A}$.}
    \vspace{-0.5cm}
   \end{minipage}\hfill
   \begin{minipage}{0.48\textwidth}
     \centering
     \tikzset{every picture/.style={scale=.8}, every node/.style={scale=.9}}
    \input{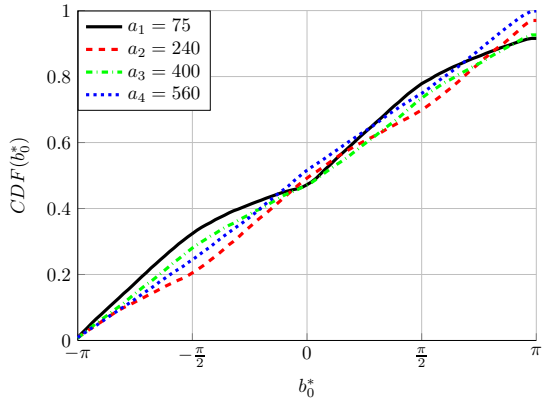}
    \subcaption{CDF of the optimal intercept $b_{0j}^*$ for all elements and users' locations, corresponding to  the four values of $a$ that constitute the set $\mathcal{A}$ in Fig. \ref{CDF} (a).}
   \end{minipage}
   \caption{ CDF plots  for $\theta_{1} = \pi$, and $\theta_{1}' = \frac{\pi}{2}$.}
\label{CDF}
\vspace{-0.5cm}
\end{figure}

To select these sets we utilize the cumulative distribution function (CDF) of the slope and the intercept obtained from \eqref{eq140}. The CDF of  ${{a}_j^*}$ in  \eqref{eq26} for all elements (i.e. $j=1,\dots, N$) and UEs' locations is shown in Fig. \ref{CDF} (a). To select the discrete set $\mathcal{A}$ with $A=\mathrm{Card}(\mathcal{A})=2^{b/2}$  slopes, we divide the CDF into $A$ equal quantiles and choose the mean of each quantile to be an element in the set $\mathcal{A}$. The case where $A = 4$ is shown in Fig. \ref{CDF} (a), with the four values of $a \in \mathcal{A}$  chosen such that  each value represents a quartile.%{$\frac{1}{2^{b/2}}$}. 

Next to find the discrete set of intercepts $b_{0}$ corresponding to each selected slope, we plot the CDF of optimal $b_{0j}^*$ in \eqref{eq107} for all elements and locations of UE as outlined in \eqref{eq140}  corresponding to  each selected value of slope. The CDF is observed to be almost linear over $[-\pi,\pi]$. An example for $A = 4$ is shown in Fig. \ref{CDF} (b), where each curve represents the CDF of $b_{0}^{*}$ over the quantile represented by slope $a_{t}$, $t=1,\dots, A$. Based on the figure, since the optimal intercept has a uniform distribution in the interval $-\pi$ and $\pi$ for each slope, we will use a uniform division of $[-\pi,\pi]$ to find the set of intercepts $\mathcal{B}_{\tilde{a}}$ corresponding to  slope $a$ with index $\tilde{a}$ in $\mathcal{A}$. The sets of intercepts across the chosen slopes are shifted by $\frac{\pi}{ \mathrm{Card}(\mathcal{B}_{\tilde{a}})}$ (to obtain triangular tiling of the (a,b) space). For example, for $\mathrm{Card}(\mathcal{B}_{\tilde{a}}) = 4$, we can use the sets $\mathcal{B}_{\tilde{a}}= \{\frac{3\pi}{4}, \frac{\pi}{4}, -\frac{\pi}{4}, -\frac{3\pi}{4}\}$ for even $\tilde{a}$, and $\mathcal{B}_{\tilde{a}}=\{\pi, \frac{\pi}{2}, 0, -\frac{\pi}{2}\}$ for odd  $\tilde{a}$. 

Once the discrete sets $\mathcal{A}$ and $\mathcal{B}_{\tilde{a}}, \tilde{a} \in \{1 ,\dots,2^{b/2}\}$ have been selected exploiting the optimal linear response in \eqref{eq25}, we need to map them to  the parameters sets for the practical $\arctan$ phase-frequency profile in \eqref{eq15}, i.e.  $\mathcal{M}$ and $\mathcal{I}_{\tilde{m}}, \tilde{m} \in \{1 ,\dots,2^{b/2}\}$, where $\mathrm{Card}(\mathcal{M}) = \mathrm{Card}(\mathcal{A})=2^{b/2}$ and $\mathrm{Card}(\mathcal{I}) = \mathrm{Card}(\mathcal{B})=2^{b/2}$. We do so by choosing the best values of the pairs $(m_t,i_{0t})$, $t=1,\dots, 2^{b/2}$ that parameterize the phase-frequency model in \eqref{eq15} corresponding to $(a_{t},b_{0t})$ that parametrizes \eqref{eq25}. The values of $m_t$  and $i_{0t}$ are chosen from  the range $m_t\in \{m_{\rm min}, m_{\rm max}\}$ and $i_{0t}\in \{ -1, 1 \}$. The pairs are selected numerically to minimize the MSE between the $\arctan$ and optimal linear phase-frequency profiles over the simulated frequency band, i.e. to minimize the MSE between the phase-frequency response in \eqref{eq15} under $(m_t,i_{0t})$ and the one in \eqref{eq25} under $(a_t,b_{0t})$. Algorithm \ref{AlgRISConfig} outlines the steps to select these sets that provide strong received signal strength in  all considered UE directions. Note that from a practical point of view, the sets $\mathcal{M}$ and $\mathcal{I}_{\tilde{m}}$ will guide the RIS unit cell design and fabrication.

%\textcolor{red}{ We need to point out that the last step of Algorithm \ref{AlgRISConfig} finds the $\arctan$ parameters sets $\mathcal{M}$ and $\mathcal{I}$ which yields the minimum MSE to the parameter sets $\mathcal{M}^*$ and $\mathcal{I}^*$ selected based on the optimized linear function in \eqref{eq25}, . This step is needed because the values of $m$ and $i_{0}$   in the practical $\arctan$ phase-frequency profile model in \eqref{eq15} that provide a good match to the optimal linear phase response in \eqref{eq25} is different from $m^*$ and $i_0^*$ due to the introduced $\arctan$ function.} 

\begin{comment}
\begin{algorithm}[t]
\caption{RIS Structure Design}
\begin{algorithmic}
\STATE 
\STATE { \textsc{ RIS Parameter Sets $\mathcal{M}$ and $\mathcal{Q}$}}
\STATE \hspace{0.5cm} Build optimal RIS phase shifts matrix $\boldsymbol{\Phi}$ using \eqref{eq21} for fixed ($\theta_{1}$,$\theta_{1}'$), and all possible values for ($\theta_{2}$, $\theta_{2}'$).
\STATE \hspace{0.5cm} Find the optimum (Intercept, Slope) for each $(\theta_{2},\theta_{2}')$ using \eqref{eq25}.
\STATE \hspace{0.5cm} Find the $PDF$ \textcolor{red}{$CMF$} of the measured slope.
\STATE \hspace{0.5cm} Find the set of slopes $\mathcal{M}$  by dividing the CDF into $M$ equal regions and choosing the mid points.
\STATE \hspace{0.5cm} Find the set $\mathcal{Q}$ representing the intercept values for  slope index $i \in \{1, \dots, card(\mathcal{M})\}$ by taking values uniformly starting from $\pi - \bmod{ (i-1,2)} \times \frac{\pi}{2 Q}$ with a step $\frac{\pi}{2 Q}$ where $Q=card(\mathcal{Q})$.

%$ as a uniform distribution starting form $\pi - \bmod{ (i-1,2)} \times \frac{\pi}{2 b}$ with a step $\frac{\pi}{2 b}$. 
\STATE \hspace{0.5cm}\textbf{return}  $\textbf{I,S}$
\end{algorithmic}
\label{AlgRISConfig}
\end{algorithm}
\end{comment}

\begin{algorithm}[t]
\caption{Phase-Frequency Profile Sets Design}
\begin{algorithmic}
\STATE 
\STATE \textbf{Initialize:} $b$, $N_{s}$, $N$, $\theta_{1}$, $\theta_{1}'$

%\STATE \textbf{Initialization:}
%\STATE \hspace{0.1cm} Set $\boldsymbol{\theta}_{2}=[\frac{\pi}{2}:0.1:\frac{3\pi}{2}]$  and $\boldsymbol{\theta}_{2}'=[\frac{\pi}{2}:0.1:\pi]$.
%\STATE \hspace{0.1cm}  Build channels $r_{k}$ in \eqref{eq19} corresponding to $\boldsymbol{\theta}_{2}$ and $\boldsymbol{\theta}_{2}'$ for  $k \in [1,N_{s}]$

\STATE \begin{enumerate}
    %\item Find $\phi_{k,i}^*$ using \eqref{eq25} for all values in $\boldsymbol{\theta}_{2}$ and $\boldsymbol{\theta}_{2}'$.
    \item Compute ${a}^*_{j}$ using \eqref{eq26} for $j \in \{1,\dots,N\}$, $\theta_{2} \in \{\frac{\pi}{2}:\Delta_{\theta}:\frac{3 \pi}{2} \}$, $\theta_{2}' \in \{0:\Delta_{\theta}:\pi\}$.
    \item Compute the $CDF$ of ${a}^*_{j}$ and divide it into $2^{b/2}$  equal quantiles.
    \item Construct the set $\mathcal{A}$ by choosing the mean of each quantile of the CDF from step (2) as $a \in \mathcal{A}$.
    \item Construct two sets of dimension $2^{b/2}$ with elements chosen equidistantly from the interval $[-\pi, \pi]$,  with the second set shifted by $\frac{\pi}{2^{b/2}}$. Use them as $\mathcal{B}_{\tilde{a}}, \tilde{a} \in \{1 ,\dots,2^{b/2}\}$ for even and odd $\tilde{a}$ respectively, where $\tilde{a}$ denotes the index of the corresponding slope $a$ in set $\mathcal{A}$.
    \item Find the $\arctan$ parameters sets $\mathcal{M}$ and $\mathcal{I}_{\tilde{m}}, \tilde{m} \in \{1, \dots, 2^{b/2}\}$ corresponding to $\mathcal{A}$ and $\mathcal{B}_{\tilde{a}}, \tilde{a} \in \{1 ,\dots,2^{b/2}\}$ such that the MSE between the proposed $\arctan$ and optimal linear phase-frequency responses over the bandwidth of interest is minimized.
    
\end{enumerate}

\STATE \textbf{Return:} $\mathcal{A}$, $\mathcal{B}_{\tilde{a}}, \tilde{a} \in \{1 ,\dots,2^{b/2}\}$, and $\mathcal{M}$,  $\mathcal{I}_{\tilde{m}}, \tilde{m}\in \{ 1,\dots,2 ^ {b/2}\}$.
\end{algorithmic}
\label{AlgRISConfig}
\end{algorithm}

% \subsubsection{ILoS received signal}\label{ILOSRIS}
% The received signal, in this case, is similar to LoS and ILoS case but with a zero LoS component.
% Similar to ILoS association, the maximum channel capacity at the receiver requires a constructive summation of the reflected signals from the RIS elements. This  condition can be written as follows:

% \begin{equation}\label{eq23}
% \begin{aligned}
%         \phi_{k,i} =   -\zeta_{k} + \nu_{k} \Tilde{d} +  2\pi \zeta \frac{f}{f_{0}} & \big(n_{y} ( \cos\theta_{1}  \sin\theta_{1}' + \cos\theta_{2}  \sin\theta_{2}' ) 
%         \\& + n_{z} ( \cos\theta_{1}' + \cos\theta_{2}' ) \big)
% \end{aligned}
% \end{equation}
% , where $ \zeta_{f}$ is the sub-carrier phase constant, which can take any value, but it must be the same for all reflected signals from RIS elements to satisfy the constructive summation condition. The optimum response in this association case is a linear response, where the intercepts will be random, and the slopes follow the same parameters in the LoS and ILoS case. 

\vspace{-.1in}
\subsection{RIS Optimization}

%Note that ; then, It will be tested for the water-filling power allocation and by adding random $\chi_{k,i}$ using Rician model.

We now proceed to solve  problem \textit{(P1)} that aims to maximize the achievable rate by designing the RIS phase-frequency profiles $\phi_{k,j}$ based on \eqref{ann} using the parameters $m_j$ and $i_{0j}$ in the sets $\mathcal{M}$ and $\mathcal{I}_{\tilde{m}}, \tilde{m} \in \{1 ,\dots,2^{b/2}\}$ respectively. We first outline the method used to optimize the RIS phase-shifts $\phi_{k,j}$ to maximize the achievable rate for a given power allocation scheme. The resulting method is then used along with water-filling power allocation that is optimal for given RIS phase-frequency profiles. Note that the RIS phase-frequency profile sets $\mathcal{M}$ and $\mathcal{I}_{\tilde{m}}$ built using Algorithm \ref{AlgRISConfig} are independent of the power allocation scheme as well as the RIS response optimization algorithm, and are selected based on the geometrical characteristics of the channel.

The optimal solution of \textit{(P1)} can be obtained using exhaustive search, which will have a prohibitively large computational complexity in the order of $O((2^b)^{N})$, since the total number of RIS configurations to choose from is $(2^b)^{N}$.  Therefore,  we propose a sub-optimal successive optimization algorithm to optimize the RIS phase shifts. We first find the initial  slope-intercept pair $({a}_j, b_{0j})$  for  RIS element $j=1,\dots, N$, denoted as $(a_j^{0},b_{0j}^{0})$  from the sets $\mathcal{A}$ and $\mathcal{B}_{\tilde{a}}$  designed in steps 3 and 4 of Algorithm \ref{AlgRISConfig}. The initial pair of values is chosen from the sets $\mathcal{A}$ and $\mathcal{B}_{\tilde{a}}$  to minimize the euclidean distance with the optimal solution ($a_{j}^{*},b_{0j}^{*}$) in \eqref{eq25} for the given UE's location. Then we find the corresponding initial slope-shift pair under the proposed $\arctan$  phase-frequency model from sets $\mathcal{M}$ and $\mathcal{I}_{\tilde{m}}$, designed in step 5 of Algorithm \ref{AlgRISConfig}. This determines the initial phase-frequency profile chosen for each RIS element and minimizes the  number of iterations needed by the proposed successive optimization algorithm to converge, as compared to using a random initialization. Then we iterate by changing the slope-intercept pair for each element one by one and updating the phase-frequency profile of that element if it increases the rate. The iterations continue until the algorithm converges in the value of the objective function in \textit{(P1)}. Algorithm \ref{AlgRISResponseWater} shows these steps while combining power allocation with RIS phase-frequency profile optimization.  Note that we update the power allocation using water-filling every time the RIS phase-frequency profile is updated for any element and check if the rate is improved. 

The convergence of Algorithm \ref{AlgRISResponseWater} is ensured by noting that the objective value of \textit{(P1)} is upper-bounded due to the constraint set in \textit{(P1)} and is non-decreasing over the iterations by applying Algorithm \ref{AlgRISResponseWater}. The algorithm has a computational complexity of $O(2^{b}\times N)$, which is linear in $N$.

\begin{algorithm}[!t]
\caption{Optimization Solution for Problem \textit{(P1)}}
\begin{algorithmic}
\STATE \textbf{Input:} $\mathcal{M}$, $\mathcal{I}_{\tilde{m}}, \tilde{m} \in \{1 ,\dots,2^{b/2}\}$, $\mathcal{A}$, $\mathcal{B}_{\tilde{a}}, \tilde{a} \in \{1 ,\dots,2^{b/2}\}$, $\theta_{1}$, $\theta_{1}'$, $\theta_{2}$, $\theta_{2}'$, $N_{s}$, $N$
\STATE \textbf{Define:} $\mathcal{S}_1=\{\mathcal{M}(1) \times \mathcal{I}_{1}, \dots, \mathcal{M}(2^{b/2}) \times \mathcal{I}_{2^{b/2}}\}$
\STATE \textbf{Define:} $\mathcal{S}_2=\{\mathcal{A}(1) \times \mathcal{B}_{1}, \dots, \mathcal{A}(2^{b/2}) \times \mathcal{B}_{2^{b/2}}\}$
\STATE \hspace{0.12cm} {Compute}  $a_{j}^{*}$ and $b^{*}_{0j}$ for $j \in \{1,\hdots,N\}$ using  \eqref{eq26} and \eqref{eq107} respectively.
\STATE \hspace{0.12cm} {Choose} ($a^{0}_{j},b^{0}_{0j}$) from the set  $\mathcal{S}_2$ that results in smallest Euclidean distance to ($a^*_{j}, b^*_{0j}$).
\STATE \hspace{0.12cm} Find the  $\arctan$ parameters ($m^{0}_{j},i^{0}_{0j}$) corresponding to  ($a^{0}_{j},b^{0}_{0j}$) from the set $\mathcal{S}_1$.

%\STATE \hspace{0.1cm} Build the sets $\mathbf{m}^0=[m^0_1, \dots, m^0_N]$ and $\mathbf{i}^0=[i^0_1, \dots, i^0_N]$.
%\STATE \hspace{0.1cm} Compute the initial phase shifts $\phi_{k,i}^o$  using  \eqref{ann} with $\mathbf{m}^0$ and $\mathbf{i}^0$.

\STATE \hspace{0.12cm} Compute the power allocation vector $\mathbf{p^{0}}$  using water-filling based on the channel in \eqref{eq19} and ($m_{j}^{0},i_{0j}^{0}$). 
\STATE \hspace{0.12cm} Set $R = 0$ and $R'$ to the achievable rate under ($m_{j}^{0}$,$i_{0j}^{0}$) and $\mathbf{p^{0}}$ using \eqref{eq13}.

   %\STATE \hspace{0.1cm}  $\mathbf{i}' = \mathbf{i}^0$\;
     %\STATE \hspace{0.1cm}   $\mathbf{m}' = \mathbf{m}^0$\;
\While{$| R - R' | > \epsilon$ }
{
$R=R'$\;

    \For{$j=1$ \KwTo $N$}
    {   
 
         \For{$u=1$ \KwTo $2^{b}$}
        {   
            $(m'_{j},i'_{0j}) = \mathcal{S}_1(u)$\;
            $\phi_{k,j}=-2 \arctan (m'_{j}(f_{k}-f_{0})+i'_{0j})$ \;
            %Find ${r}_k$, $k=1,\dots, N_s$, using \eqref{eq19}\;
            Compute the water-filling power allocation $\mathbf{p}'$ based on \eqref{eq19} and ($m'_{j},i'_{0j}$). \;
            Compute $R''$ using \eqref{eq13} \;
            \If{$R'' > R'$}
            {
                %$\mathbf{i}^* = \mathbf{i}'$\;
                %$\mathbf{m}^* = \mathbf{m}'$\;
                %$\mathbf{p}^* = \mathbf{p}'$\;
                ($m_{j},i_{0j}$ )= ($m_{j}',i_{0j}'$)\;
                $\mathbf{p}=\mathbf{p}'$\;
                $R' = R''$\;
            }
        }
    }
}
   % \STATE \textbf{Return:}  $\mathbf{I}^*$, $\mathbf{M}^*$, $\mathbf{P}^*$
\STATE \textbf{Return:}  (${m}_j$, ${i}_{0j}$) $j=1,\dots, N$, $\mathbf{p}$
\STATE $\phi^*_{k,j} =-2 \arctan (m_{j}(f_{k}-f_{0})+i_{0j})$, $j=1,\dots, N$, $k=1,\dots, N_s$
\end{algorithmic}
\label{AlgRISResponseWater}
\end{algorithm}

\vspace{-.1in}
\section{Numerical Results} \label{resultSection}
In this section, we validate the proposed RIS design by conducting simulations for an RIS-assisted OFDM system, where the achievable rate is plotted as a function of UE's position. The parameter values are set as: Tx position is $(0,0,3)$ (in metres), RIS position is $(100,0,3)$, UE position is on a circle centred at the RIS with radius $r = 15$, $\Delta = \frac{\lambda_{0}}{2}$,  $f_{0} = 2.5$ GHz,  $\Delta_{f} = 200$ KHz, $P = 1$ dBm, and $\sigma^{2} = -174$ dBm. The antenna parameters for the Tx (UE) are set as $\theta_{\rm Tx} = \frac{\pi}{6}$ ($\theta_{\rm UE} = \frac{\pi}{4}$) and $\phi_{\rm Tx} = \frac{\pi}{4}$ ($\phi_{\rm UE} = \frac{\pi}{2}$). In the first few results, we consider \textit{equal power allocation} among OFDM sub-carriers while performing RIS phase-frequency profile optimization, and then incorporate joint power allocation and RIS phase-frequency profile optimization in the subsequent results. Moreover, to focus on the impact of the RIS, the first few results will be for the case where the direct link between Tx and UE is blocked. Then we will incorporate the direct link in the results and verify the insights for this more general model considered in the analysis. 

In all figures, we plot the performance under the proposed $\arctan$ phase-frequency profile, as well as under a constant (flat) phase-frequency profile with maximum (i.e. unity) amplitude reflection coefficient. For the flat phase-frequency profile, we consider that unit cell $j$ provides a constant phase shift from the set $\{-\pi, -\pi +\frac{2 \pi}{2^{b}}, \dots, \pi-\frac{2 \pi}{2^{b}} \}$ over all frequencies. 

%Note that the frequency flat phase response model that assumes zero slope and different intercepts has been widely used in the RIS literature and is not a practical model when it comes to wide-band systems due to the  frequency dependency of the phase response.

%Finally, we show how the system capacity changes for Rician channel.

%It is worth mentioning that the BS has a more directive antenna compared with the receivers' antenna, this assumption is used in several works in mm waves and THz systems \cite{APHPBWNour}, where the antenna of BS has high gain. In addition, in practical scenarios, the RIS is beamforming the signal toward a UE in the azimuth plane, so it is a practical assumption that the RIS has $N_{y} \geq 10 \times N_{z}$. 

%It is worth mentioning that, in this work, we are examining a channel with two components; the LoS component $\mathbf{h}_d$ and the one through the RIS $\mathbf{g}$. The UE in our work will be closer to the RIS than the BS, which helps demonstrate the effects of the RIS more effectively. For instance, in the case where the distances between the BS and the RIS and the RIS and the UE are equal (i.e., $d_{1} = d_{2}$), the FSPL component of the LOS signal will be proportional to $\frac{1}{d_{d}}$, while the FSPL of the received signal from the RIS will be proportional to $\frac{1}{d_{d}^2}$.

\vspace{-.1in}
\subsection{Impact of  $b$ and $N$}
%For example, when $b = 2$, using $i_{0} = [-\frac{\pi}{2}, \frac{\pi}{2}]$ for a fixed slope provides the same results when using different slopes for the same $i{0}$. 

\begin{figure}[!t]
      \tikzset{every picture/.style={scale=.8}, every node/.style={scale=.9}}
        \input{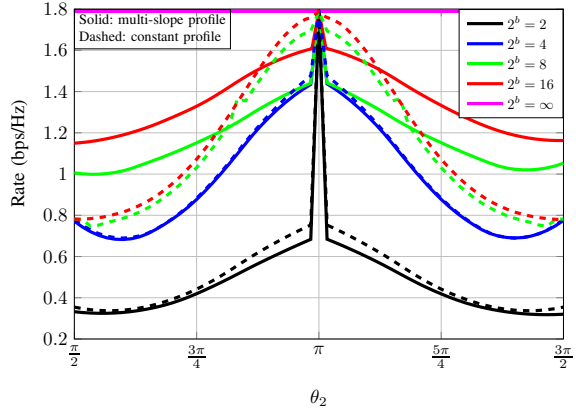}
        \caption{Data rate for different values of $b$.  }
        \label{NbitsIloS}
        \vspace{-0.2cm}
   \end{figure}

Fig. \ref{NbitsIloS} shows the achievable rate for $N_{s} = 256$, $N_{y} = 200$, $N_{z} = 1$, and various values of $b$ that determines the total number of RIS phase-frequency profiles that can be selected at each unit cell. As expected, the achievable rate increases as  $b$ increases for both models. However, the improvement in the achievable rate for the constant phase-frequency model becomes insignificant for $b > 3$, because we are essentially adding more intercept values for the same phase-frequency slope leading to a saturated performance. On the other hand, the $\arctan$  model allows for multiple slopes in the set $\mathcal{M}$ and the intercept points are spread over multiple values of slopes, so increasing the number of possible responses will result in a finer sampling of the slope-intercept space which results in higher achievable rate. When the number of available states is low (i.e. $b=1,2$), the performance of the multiple slope $\arctan$ model is similar to the constant phase-frequency model because in both cases, the sampling of the slope-intercept space is sparse, leading to poor performance. We note that the performance of the constant phase-frequency profile is better than the multiple-slope $\arctan$ phase-frequency profile near the specular angle (i.e. where $\theta_1=\theta_2$). This result is expected as $\small( \cos\theta_{1} \sin\theta_{1}' + \cos\theta_{2} \sin\theta_{2}' \small) $ and $\small( \cos\theta_{1}' + \cos\theta_{2}' \small) $   in \eqref{eq26} are minimized at this angle, leading to a small required slope and making the zero slope solution closest to the optimum solution. The case where $2^{b} = \infty$ corresponds to the case where the phase shifts can be chosen independently of the frequency from an infinite number of states, i.e., at each frequency an independent phase shift can be chosen (similar to the dashed response in Fig.~\ref{SubSetResponse}). This acts as a performance upper bound for the considered system model, and we note that the $\arctan$ phase-frequency profile yields a closer performance to this benchmark than the constant phase-frequency profile over a wider range of user directions. 
   \begin{figure}[!t]
     \centering
        \tikzset{every picture/.style={scale=.8}, every node/.style={scale=.9}}
    \input{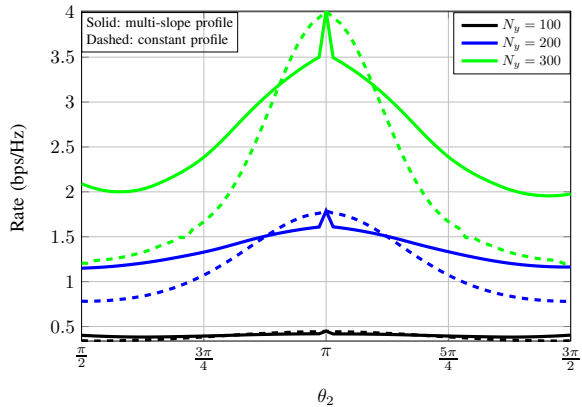}
    \caption{Data rate  for different values of $N=N_y$.}
    \label{NElementsILoS}
    \vspace{-0.2cm}
\end{figure}

Fig. \ref{NElementsILoS} compares the achievable rate for the two phase-frequency profile models for $N_{s} = 256$, $b = 4$, $N_{z} = 1$, and a varying  number of RIS elements $N=N_{y}$. We observe that as the number of elements increases, the advantage of using different slopes under the proposed $\arctan$ phase-frequency model becomes more noticeable. This can be explained by noting that that as $n_{z,j}$ or $n_{y,j}$ increases (with an increase in $N$), the optimal slope in \eqref{eq26} also increases, leading to the need for higher slopes for some elements in the proposed $\arctan$ phase-frequency response function in \eqref{eq15}. Consequently, the proposed model, which allows for different values of the slope for different elements,  performs better, particularly for UE positions away from the specular angle. 

%It's worth noting that changing the element spacing $\Delta$ has a similar effect as increasing the number of elements at the RIS $\mathrm{N}$. %however, increasing $\Delta$ increases the level of minor lobes of the RIS, making the proposed model invalid, so its effect was not studied. 

To measure the extended coverage provided by the proposed model, we evaluate the area over which a given QoS constraint is satisfied for both phase-frequency models. Specifically, we evaluate the percentage of angles of arrival $\theta_{2}$ corresponding to different UE locations, where the corresponding achievable rate is higher than a predefined threshold  $R_{th}$, and show the results in Table \ref{NelementPercentage}. Interestingly, the $\arctan$ phase-frequency profile model provides competitive performance for small values of $R_{th}$, and outperforms the constant phase-frequency profile model for high values of $R_{th}$ as well as for large values of total number of possible phase-frequency configurations $2^{b}$.

% our work, we will use high-$QOS$ thresholds to compare channel capacity between the constant and variable slope configurations of the RIS, at different UE locations. Additionally, we will determine the lowest system capacity using minimum-rate measurements. Table \ref{NelementPercentage} displays the range of angles where the channel capacity exceeds the high-$QOS$ thresholds $R_{\mathrm{th}}$ for $\arctan$  phase profile, compared to the constant phase profile, with different number of states, and a constant $\mathrm{N} = 200$.

\begin{table}
\begin{center}
\caption{Percentage of UE's locations where $R\geq R_{th}$ under $\arctan$ and constant phase-frequency profiles.} % in ILoS/ILoS + LoS association.}
\label{NelementPercentage}
\begin{tabular}{| c | c | c |}
\hline
$2^b|R_{th}$ & $\arctan$ phase profile & Const phase profile\\
\hline
$2|0.4$ & 54\% & 57 \% \\
\hline
$4|0.8$& 57\% & 57 \% \\
\hline
$8|1$& 97\% & 51 \% \\
\hline
$16|1.2$&75\% & 41 \%\\
\hline 
\end{tabular}
\end{center}
\vspace{-.2in}
\end{table}

\vspace{-.1in}

\subsection{Impact of Tx-RIS Link} \label{changetheta}

%When $\theta_{1} = \frac{\pi}{2}$, a symmetrical response is observed around $\theta_{1}$, as the value of $\cos\theta_{2}$ remains consistent on both the left and right sides. However, for other values of $\theta_{1}$, a similar response is observed with a shift of the specular angle corresponding to the value of $\theta_{1}$. 

\begin{figure}[!t]
     \centering
        \tikzset{every picture/.style={scale=.78}, every node/.style={scale=.9}}
    \input{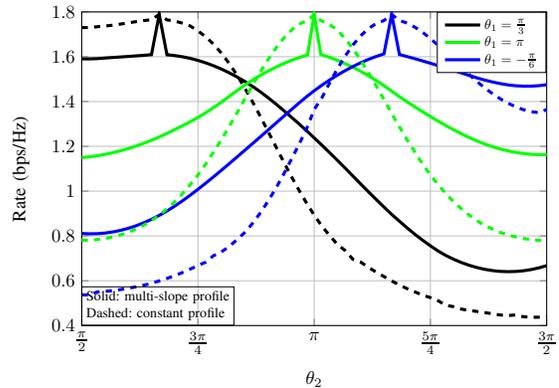}
    \caption{Data rate for different $\theta_{1}$ values as a function of $\theta_{2}$.}
    \label{theta1}
    \vspace{-0.5cm}
   \end{figure}

Fig. \ref{theta1} illustrates the achievable rate for $N_{s} = 256$, $N_{y} = 200$, $N_{z} = 1$, $b = 4$, and different values of the angle of departure $\theta_{1}$ (from the Tx to the RIS). Changing the value of $\theta_{1}$ will result in a different CDF of the optimal slope and, therefore, will result in a different set of slopes $\mathcal{A}$  under the optimal linear phase-frequency solution in \eqref{eq25}, which is then mapped to $\mathcal{M}$ under the proposed $\arctan$ phase-frequency profile. For each value of $\theta_{1}$ we found the  sets $\mathcal{M}$ and $\mathcal{I}_{\tilde{m}}, \tilde{m} \in \{1 ,\dots,2^{b/2}\}$ using Algorithm \ref{AlgRISConfig}. We then used the selected sets in Problem \textit{(P1)} and solved it using Algorithm \ref{AlgRISResponseWater}. In a practical deployment, the locations of the Tx and RIS are fixed, so the RIS configuration is established based on a known, fixed value of $\theta_{1}$, and we do not need to regenerate the sets $\mathcal{M}$ and $\mathcal{I}_{\tilde{m}}$, and re-fabricate the RIS. We see a similar performance trend across $\theta_2$ for all values of $\theta_1$ with the specular angle shifted according to $\theta_2=\theta_1$.

Table \ref{ThetaBS} shows the QoS performance for both phase-frequency profile models, where the values are tabulated for different values of $\theta_{1}$ and  $R_{th} = 1.2$. As $\theta_{1}$ approaches zero, the percentage of UE's locations where  $\arctan$ phase-frequency response provides the required QoS increases because the terms $\small( \cos\theta_{1} \sin\phi_{1} + \cos\theta_{2} \sin\phi_{2} \small) $ and $\small( \cos\phi_{1} + \cos\phi_{2} \small) $ in $\Delta_{y,j}$ and $\Delta_{z,j}$ in \eqref{eq26} become significant, implying that the multiple slopes solution for different elements is better than the constant phase-frequency profile solution.

   \begin{figure}[t]
     \centering
     \tikzset{every picture/.style={scale=.78}, every node/.style={scale=.9}}
    \input{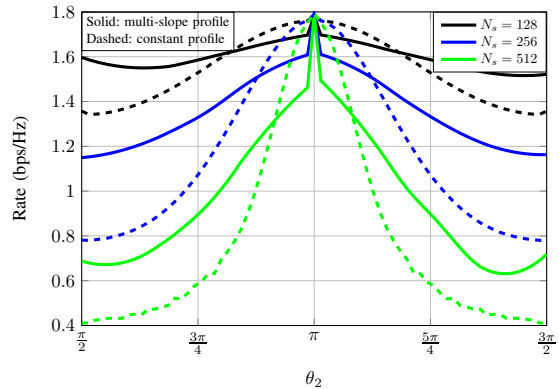}
    \caption{Data rate  against $\theta_{2}$ for different values of $N_s$.}
    \label{NsubcarriersILoS}
    \vspace{-0.5cm}
\end{figure}

\begin{table}
\begin{center}
\caption{Percentage of UE's locations where $R\geq R_{th} = 1.2$ under $\arctan$ and constant phase models for different $\theta_{1}$.}
\label{ThetaBS}
\begin{tabular}{| c | c | c | c |c |c |}
\hline
Profile & $\theta_{1} = 0$& $\theta_{1} = \frac{\pi}{6}$ & $\theta_{1} = \frac{\pi}{4}$ & $\theta_{1} = \frac{\pi}{3}$ & $\theta_{1} = \pi$\\
\hline
$\arctan$& 75 \%& 64 \%& 58 \%& 52 \%& 50 \% \\
\hline
Const& 41 \%& 54 \%& 50 \%& 42 \%& 34 \%\\ 
\hline 
\end{tabular}
\end{center}
\vspace{-.1in}
\end{table}

\vspace{-.1in}
\subsection{Impact of Number of Sub-carriers}
Next we plot the achievable rate against $N_s$  in Fig. \ref{NsubcarriersILoS} for $N_{y} = 200$, $N_{z} = 1$, and $b = 4$. For all values of $N_{s}$, the rate exhibits a similar trend across $\theta_{2}$, with the constant phase-frequency profile performing better near the specular region and the $\arctan$ model providing better performance elsewhere. However, as the number of sub-carriers increases, the angular band where the rate achieved under the $\arctan$ model is better than that achieved under the constant phase-frequency model increases. This can be explained using  \eqref{eq25}, where as $N_s$ increases, the difference between the constant phase-frequency response and the optimal phase-frequency response increases, particularly at the edge sub-carriers. This difference is less significant under the $\arctan$ model, where a proper choice of ($m_j,i_{0j}$) for each element $j$ provides a close match to the optimal response (which also varies with frequency). 

%On the other hand, when the number of sub-carriers is low, the phase change over a narrow band is minimal, leading to a smaller difference between the constant phase-frequency model and the $\arctan$ phase-frequency model. 

%Despite the correction provided by using different slopes in the $\arctan$ model, there is a loss caused by phase saturation at $\pi$ and $-\pi$, which limits the range of possible values for $m$ and the distribution of the intercept points $i_{0}$. As $m$ becomes larger, the corresponding $i_{0}$ is restricted to a narrower range and saturation occurs for larger number of sub-carriers (especially when $i_{0}$ is closer to $  \pi $ or $-\pi$). 

Table \ref{NsubcarrierTable} shows the percentage of UE locations where the achievable rate $R_{a}$ of the $\arctan$ phase-frequency model exceeds the achievable rate $R_{c}$ of the constant phase-frequency profile  model, and the maximum rate gain defined as $\underset{\theta_{2}}{\max} \frac{R_{a} - R_{c}}{R_{c}}$, for $N = 200$ and $N = 300$ and different values of $N_{s}$. The tabulated results show that as $N$ and $N_s$ increase, the percentage of UE's location where $R_{a} > R_{c}$ as well as the rate gain increases.  These results show that the proposed multiple-slope $\arctan$ phase-frequency model in \eqref{eq15}  is not only practical but it also improves the achievable rate in a wide-band OFDM system, where the dependence of the RIS phase response on frequency should not be ignored. 

%and the analytical discussion provided in \ref{RISPara}, where increasing $N_{s}$ or $N$ shows the added value of using multiple slope $\arctan$ frequency-phase profile better.

\begin{table}
\begin{center}
\caption{Percentage of UE's location where  $\arctan$ phase model provides better rate than  constant phase model as well as the maximum percentage rate gain of $\arctan$  model.}
\label{NsubcarrierTable}
\begin{tabular}{| c | c | c | c |c |}
\hline
 & \multicolumn{2}{c|}{$N = 200$} & \multicolumn{2}{c|}{$N = 300$}  \\
\hline
$N_{s}$ & $R_{a} > R_{c}$ & $\underset{\theta_{2}}{\max} {\frac{R_{a} > R_{c}}{R_{c}}}$ & $R_{a} > R_{c}$ & $\underset{\theta_{2}}{\max} {\frac{R_{a} > R_{c}}{R_{c}}}$  \\
\hline
 $128$ & 64 \%& 18 \%& 68  \%& 37 \% \\
\hline
$256 $& 76 \%& 49 \%& 81 \%& 73 \% \\
\hline
$512 $& 86 \%& 74 \%& 93 \%& 91 \% \\
\hline 
\end{tabular}
\end{center}
\vspace{-.1in}
\end{table}

%As these channels follow a random distribution, the power allocation will be unpredictable. We can't use \ref{AlgRISResponse} to optimize the RIS response in the water-filling power allocation, as the power allocation result must be considered while choosing the optimum RIS response allocation.

\vspace{-.1in}
\subsection{Changing Power Allocation Algorithm}
\begin{figure}[!t]
     \centering
    \tikzset{every picture/.style={scale=.77}, every node/.style={scale=.9}}
    \input{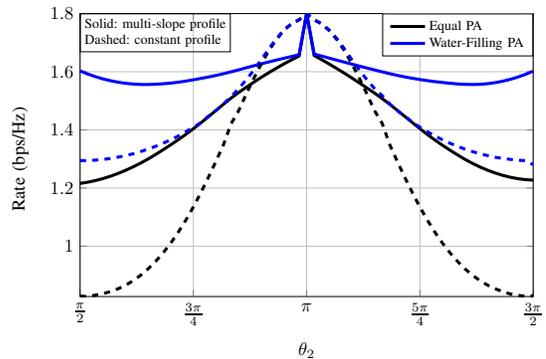}
    \caption{Impact of power allocation on data rate.}
    \label{Water_equalPA}
   \end{figure}

Previously we focused only on the impact of the  $\arctan$ phase-frequency profile model in \eqref{eq15} and its optimization on the achievable rate in wide-band systems, and therefore considered equal power allocation across the sub-carriers. Next, in Fig. \ref{Water_equalPA}, we show the achievable rate under the joint optimization of power allocation and RIS phase-frequency response as outlined in Algorithm \ref{AlgRISResponseWater} for $N_{s} = 256$, $N_{y} = 200$, $N_{z} = 1$, and $b = 4$. We also show the achievable rate under RIS phase-frequency response optimization with equal power allocation. As expected, the joint optimization of power allocation and RIS phase-frequency response results in better performance, with the $\arctan$ phase-frequency profile providing higher rates for values of $\theta_2$ that are away from the specular angle.

% This is because under equal power allocation algorithms and a constant slope RIS configuration, equal power was assigned to both mid-band carriers (where their responses are close to the optimal ones) and edge-band carriers (where there is a significant difference between the provided responses and the optimal ones). In contrast, using the water-filling algorithm, less power is assigned to edge-band carriers and more power to mid-band carriers, which results in better achievable rate.

\vspace{-.1in}
\subsection{Impact of Direct Tx-UE Link}
\begin{figure}[!t]
     \centering
     \tikzset{every picture/.style={scale=.77}, every node/.style={scale=.9}}
    \input{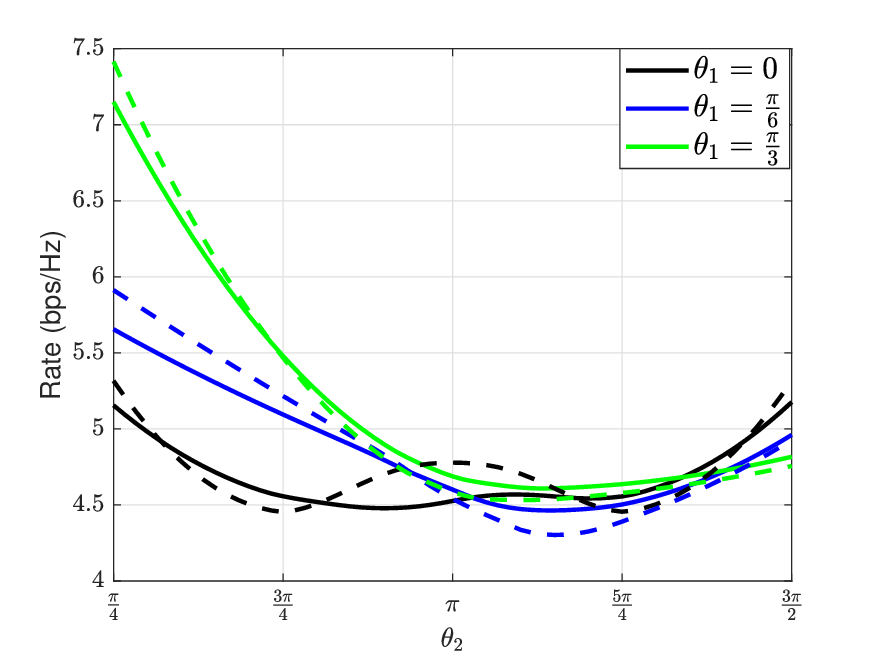}
    \caption{Data rate against $\theta_2$ for different $\theta_{1}$ values considering both direct and RIS-assisted links.}
    \label{LoSILoS}
    \vspace{-0.5cm}
\end{figure}

In this section, we study the effect of having a direct link between the Tx and the UE, as well as a reflected link through the RIS. We find that including the direct channel does not affect the slope  ${a^*_{j}}$ in \eqref{eq26} of the optimal response of each RIS element because we can use Tx precoding over all sub-carriers to cancel its effect, i.e. cancel the effect of $d_d$ as discussed after   \eqref{eq21}. 
%As a result, the output of the RIS configuration algorithm (Algorithm \ref{AlgRISConfig}) will be the same regardless of the presence of the direct link. However, the result of the RIS response optimization algorithm (Algorithm \ref{AlgRISResponseWater}) will be different, as $d_{d}$ in equation \eqref{eq107} will impact the choice of phase shift at different RIS unit cells. 
Fig. \ref{LoSILoS} shows the achievable rate for $N_{s} = 256$, $N_{y} = 200$, $N_{z} = 1$, $b = 4$,  Tx-RIS distance of $2$ km, and different angles of departure $\theta_{1}$. The behaviour of the rate is similar to that observed earlier. However, the range of angles $\theta_{2}$ where the $\arctan$ phase-frequency profile performs better than the constant phase-frequency profile is shifted with less gain margin. The gain now appears smaller because the direct channel is stronger than the reflected channel though the RIS. Larger gains are expected when $N$ is larger.

%It is worth noting that the same path loss coefficient was used for both links (the direct and reflected links), so a significant distance between the RIS and Tx is needed to demonstrate the effect of the reflected signal from the RIS. This assumption is valid as the primary purpose of using RIS is to provide a  controllable link in the absence of a direct signal, which is expected to be in the far region from the Tx. 

   \vspace{-0.5cm}

\subsection{Performance under Rician Channels}

\begin{figure}[!t]
     \centering
  \tikzset{every picture/.style={scale=.77}, every node/.style={scale=.9}}
    \input{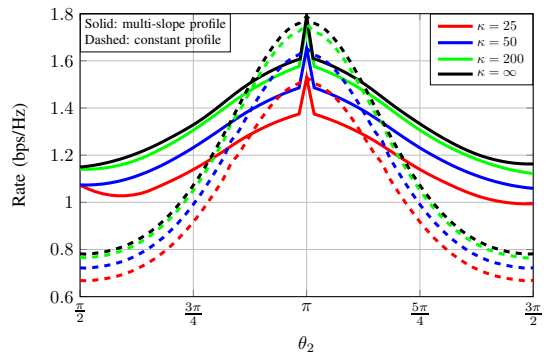}
    \caption{Data rate for different Rician factors $\kappa$.}
    \label{RicianChannel}
    \vspace{-0.5cm}
   \end{figure}

The results so far consider the channels to be LoS dominated as we assumed the NLoS component $\tilde{\chi}_{k,j}$ in \eqref{eq19} to be zero. In this section, both the Tx-RIS and RIS-UE channels will follow Rician fading models that account for both LoS and NLoS channel components. The channel coefficient in  \eqref{eq5} and \eqref{eq27}  will be modified to account for the Rician factors as 
\cite{RicianChannel}: \vspace{-.1in}
\begin{align}
        \label{eq24}
    &h_{k,1,j} = \sqrt{\frac{\kappa}{\kappa+1}} \alpha_{k,1}  e^{-\frac{j2\pi f_{k}}{c} \Tilde{d}_{1,j}} + \chi_{k,1,j}, \\
        \label{eq108}
    &h_{k,2,j} = \sqrt{\frac{\kappa}{\kappa+1}} \alpha_{k,2}  e^{-\frac{j2\pi f_{k}}{c} \Tilde{d}_{2,j}} + \chi_{k,2,j}, 
   \end{align}
where $\chi_{k,1,j} = \sqrt{\frac{1}{\kappa+1}}{\alpha_{k,1}} l_{k,1,j}$,   $\chi_{k,2,j} =  \sqrt{\frac{1}{\kappa+1}}{\alpha_{k,2}} l_{k,2,j}$, $\kappa$ is the Rician factor, and $l_{k,1,j}, l_{k,2,j} \sim CN(0,\sigma^2)$ are complex Gaussian random variables with zero mean and $\sigma^2$ variance. Despite the randomness of $\chi_{k,1,j}$  and $\chi_{k,2,j}$, it is impractical to adjust the configuration of the RIS elements continuously whenever $\chi_{k,1,j}$ and $\chi_{k,2,j}$ change. Therefore, we will continue to use Algorithms \ref{AlgRISConfig} and \ref{AlgRISResponseWater} to select the RIS parameter sets $\mathcal{M}$ and $\mathcal{I}_{\tilde{m}}$ and optimize the phase-frequency profiles based on the location of the UE. Fig. \ref{RicianChannel} illustrates the effect of  Rician fading on achievable rate for $N_{s} = 256$, $N_{y} = 200$, $N_{z} = 1$, $b = 4$, and different values of $\kappa$. When $\kappa = 200$, the rate decreases by approximately $2\%$ compared to the purely LoS scenario, which is expected due to the additional random phase rotations introduced by the Rician channel, which were not accounted for in the RIS design. As $\kappa$ decreases, the rate also decreases as the channel becomes more random (weaker line-of-sight), and the RIS is unable to compensate for the effects of random fading. It is worth noting that both the constant and the $\arctan$ phase-frequency profile models experience a similar level of degradation.

\vspace{-.05in}
\section{Extension to Multi-UE MISO Scenario}\label{SecMISO}

This section provides an extension of the proposed  RIS phase-frequency profile selection and optimization methods to the multi-UE MISO scenario. We first describe the system model,  and then  solve the sum-rate maximization problem.

\vspace{-.1in}
\subsection{Signal Model and Problem Formulation}

We consider a multi-UE MISO communication scenario, where the Tx has a rectangular antenna array with $Q = Q_{x}Q_{z}$ elements, where $Q_{x}$  and $Q_z$ are  the numbers of antennas along the x-axis and z-axis respectively, arranged with an antenna spacing of $\Delta_{T}$. The position of antenna $j$ is  $[\Delta_{T} q_{x,j},0,z_{b} + \Delta_{T} q_{z,j}]$, where $q_{x,j}$ and $q_{z,j}$ can be written  as $\lfloor \frac{j-1}{Q_{y}} \rfloor+1-Q_{z}/2$ and $\mod (j-1,Q_{y}) + 1 - Q_{x}/2$, respectively. Moreover, we consider $L$ UEs in the system. The RIS and UEs have the same configuration described in Sec.~\ref{systemModel}. 

The received base-band OFDM signal at  UE $l$  corresponding to sub-carrier $k$ can be written as \cite{MISORIS}

\begin{equation}
    y_{k,l} = \big (\mathbf{h}_{k,d,l} +  \mathbf{g}_{k,l} \big )^H \mathbf{u}_{k,l} \sqrt{p_{k,l}}x_{k,l} + z_{k,l},  
   \label{eq103}
\end{equation}
where  $\mathbf{h}_{k,d,l}$ is the Tx-$\text{UE}_{l}$ direct channel, $\mathbf{g}_{k,l} $ is the channel from the Tx to  $\text{UE}_{l}$ through the RIS,  $\mathbf{u}_{k,l}$ is the precoding vector applied at the Tx to $\text{UE}_{l}$'s data signal $x_{k,l}$, $p_{k,l}$ is the power allocated to $\text{UE}_{l}$'s signal, and $z_{k,l} \in \mathbb{C}$ is the AWGN at $\text{UE}_{l}$.  The channel models for each element of $\mathbf{h}_{k,d,l}$ and $\mathbf{g}_{k,l} $ are as provided in Sec.~\ref{systemModel}. Since our main focus is on the effect of RIS phase-frequency profile modelling and optimization,  optimizing the precoding vectors is out of the scope of this work. Thus, we will  use the well-known zero-forcing (ZF) \cite{ZeroForcing} and  maximum ratio transmission (MRT) \cite{MaxRatTRan} precoding methods to implement $\mathbf{u}_{k,l}$. The sum-rate maximization problem for this scenario is formulated next.
\vspace{-.1in}
\begin{subequations}
\label{eq104}
 \begin{alignat}{2} \textit{(P2)} \hspace{.15in}
&\!\max_{\begin{subarray}{c}
  \boldsymbol{\Phi}, 
  \mathbf{P}
  \end{subarray}} \hspace{.1in}         &\hspace{.1in} & \frac{1}{N_{s}} \sum_{l=1}^{L} \sum_{k=1}^{N_{s}} \log_{2} \small(1+ \zeta_{k,l} \small) \label{obj2}\\
&\text{subject to} &      & \mathbb{E}[\lVert \mathbf{x}\rVert^2] = tr(\mathbf{P U^H U}) \leq P, \\
&&      & \phi_{k,j} =-2\arctan(m_{j}(f_k-f_{0}) +i_{0j}), \nonumber  \\
&& & \hspace{.04in} \text{for } j=1,\dots, N, k=1,\dots, N_s \\
& & & m_{j}\in \mathcal{M}, \\
& & & i_{0j}\in \mathcal{I}_{\tilde{m}}, \tilde{m}=\text{index}(m_j)= \in \{1 ,\dots,2^{b/2}\}
\end{alignat}
\end{subequations}
where $U = [\mathbf{u}_{1},\mathbf{u}_{2}, \hdots,\mathbf{u}_{L}] \in \mathbb{C}^{Q \times L}$ is the precoding matrix, $\mathbf{P}=\text{diag}(p_{1,1}, \dots, p_{N_{s},L})$ is the power allocation matrix, and $\zeta_{k,l}$ is the signal-to-interference-plus-noise ratio (SINR) at UE $l$ corresponding to sub-carrier $k$ which is given as 
\begin{equation}
\label{eq105}
    \zeta_{k,l} = \frac{ 
 \lVert (\mathbf{h}_{k,d,l} + \mathbf{g}_{k,l})^H \mathbf{u}_{k,l} \rVert^2 p_{k,l}}{\sum_{j \neq l} \lVert (\mathbf{h}_{k,d,l} + \mathbf{g}_{k,l})^H \mathbf{u}_{k,j} \rVert^2 p_{k,j} +  \sigma^{2} \Delta_f}
\end{equation}

\vspace{-0.2in}
\subsection{Problem Solution}
Similar to the SISO scenario, we will first study the selection of the RIS phase-frequency profiles parameter sets $\mathcal{M}$ and $\mathcal{I}_{\tilde{m}}$ that is favourable for sum rate maximization in a multi-UE MISO scenario. Then we solve  Problem \textit{(P2)} based on the designed sets. We design $\mathcal{M}$ and $\mathcal{I}_{\tilde{m}}$ using the same procedure as in Algorithm \ref{AlgRISConfig}, by decomposing the multi-UE MISO system into  multiple single-UE SISO systems by considering each Tx-antenna-UE pair as a single-UE SISO system\footnote{This is done for the purpose of selecting $\mathcal{M}$ and $\mathcal{I}_{\tilde{m}}, \tilde{m} \in \{1 ,\dots,2^{b/2}\}$ only, but not for the optimization of (P2).}. With this consideration, Algorithm \ref{AlgRISConfig} can be straightforwardly generalized to the multi-UE MISO case by replacing the CDF  of optimal slope corresponding to the single UE's locations in step (2) of Algorithm \ref{AlgRISConfig} with the CDF corresponding to all Tx antenna elements $q=\{1,\dots, Q\}$ and all locations of UEs $l=\{1,\dots, L \}$. Specifically, in step (1), we will find the optimal slope $a_{j,q,l}^*$  with respect to each Tx antenna $q=\{1,\dots, Q\}$ and UE $l=\{1,\dots, L\}$ for all values of $\theta_{2}$ and $\theta_{2}'$. The rest of the steps are the same and follow by plotting the CDF of $a_{j,q,l}^*$, $\forall j,q,l$. Note that the extended Algorithm \ref{AlgRISConfig} only needs geometrical information related to the locations of Tx and RIS and the set of possible locations of users.

The RIS response optimization algorithm to solve \textit{(P2)} based on the designed sets  $\mathcal{M}$ and $\mathcal{I}_{\tilde{m}}, \tilde{m} \in \{1 ,\dots,2^{b/2}\}$ is similar to Algorithm \ref{AlgRISResponseWater}, with the following additional steps. For each element $j$, when we test a new state $u=\{1,\dots, 2^b\}$ based on  $\mathcal{M}$ and $\mathcal{I}_{\tilde{m}}$ resulting in new ${\phi}_{k,j}$, we will recompute the power allocation vector $\mathbf{p}=[p_{1,1}, \dots, p_{N_{s},L}]$, and the precoding matrix $\mathbf{U}$ (based on ZF or MRT) before computing the new sum-rate in \eqref{obj2}, and check if the sum-rate improves. Note that the extended Algorithm \ref{AlgRISResponseWater} only relies on knowledge of users' locations. The performance of the extended algorithm for the multi-UE MISO case is evaluated next.  
%It should be noted that while the RIS configuration and response optimization are done based on the knowledge of UE locations only, the pre-coding matrix design requires channel state information (CSI) at the base station. 

% Firstly, we study the cases where the number of available states at the RIS ($2^b$) and the number of RIS elements ($N$) are changing. Secondly, we examine the impact of different angles of departure from the Tx to the RIS ($\theta_{2}$). Next, we show the achievable rate behaviour for different system frequency band, and different power allocation algorithm. Finally, we show system performance in the case of Rician channel.

\vspace{-.1in}
\subsection{Numerical Results}
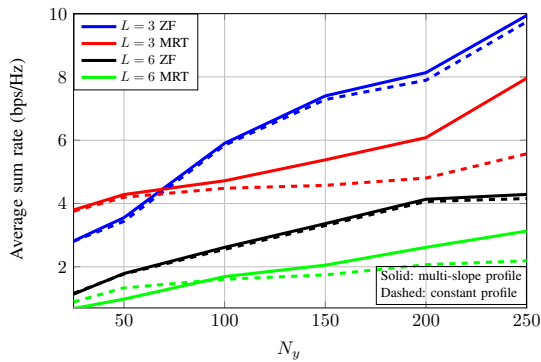
\begin{figure}[!t]
     \centering
    \tikzset{every picture/.style={scale=.77}, every node/.style={scale=.9}}
\begin{tikzpicture}

\begin{axis}[%
width=4in,
height=2.6in,
at={(0.758in,0.488in)},
scale only axis,
xmin=25,
xmax=250,
xlabel style={at={(axis cs:132,0.7)},anchor=north},
xlabel={$N_{y}$},
xtick={50,100,150,200,250},
xticklabels={{$50$},{$100$},{$150$},{$200$},{$250$}},
ymin=0.7,
ymax=10,
ylabel style={at={(axis cs:22,5.4)},anchor=north},
ylabel={Average sum rate (bps/Hz)},
axis background/.style={fill=white},
xmajorgrids,
ymajorgrids,
legend style={{at =  (axis cs:25,10)}, nodes={scale=0.75, transform shape}, anchor=north west, legend cell align=left, align=left, draw=black}
]

\addplot [color=blue, line width=1.5pt]
  table[row sep=crcr]{%
25	2.81541058170648\\
50	3.55155909012307\\
100	5.90235317667356\\
150	7.40023730357496\\
200	8.13517687711598\\
250	9.94059876822103\\
};
\addlegendentry{$L = 3$ ZF}

\addplot [color=red, line width=1.5pt]
  table[row sep=crcr]{%
25	3.79514248546474\\
50	4.28333333333333\\
100	4.71849830969956\\
150	5.37987064510421\\
200	6.08166666666667\\
250	7.95816666666667\\
};
\addlegendentry{$L = 3$ MRT}

\addplot [color=black, line width=1.5pt]
  table[row sep=crcr]{%
25	1.13536697976626\\
50	1.78759505068567\\
100	2.61337311729433\\
150	3.3640640758257\\
200	4.13370962797444\\
250	4.28655584134885\\
};
\addlegendentry{$L = 6$ ZF}

\addplot [color=green, line width=1.5pt]
  table[row sep=crcr]{%
25	0.680607711608397\\
50	0.97818353282575\\
100	1.69085\\
150	2.05208333333333\\
200	2.61459016666667\\
250	3.12757441666667\\
};
\addlegendentry{$L = 6$ MRT}

\addplot [color=blue, dashed, line width=1.5pt, forget plot]
  table[row sep=crcr]{%
25	2.80699845266797\\
50	3.43587714291063\\
100	5.84321135275357\\
150	7.28461821302186\\
200	7.89420268438635\\
250	9.74082493542915\\
};
\addplot [color=red, dashed, line width=1.5pt, forget plot]
  table[row sep=crcr]{%
25	3.7571910606101\\
50	4.19766666666667\\
100	4.48257339421458\\
150	4.57289004833858\\
200	4.80451666666667\\
250	5.57071666666667\\
};
\addplot [color=black, dashed, line width=1.5pt, forget plot]
  table[row sep=crcr]{%
25	1.16260561014135\\
50	1.77001504464586\\
100	2.55192291310175\\
150	3.30274407822135\\
200	4.06376822256341\\
250	4.15662945584968\\
};
\addplot [color=green, dashed, line width=1.5pt, forget plot]
  table[row sep=crcr]{%
25	0.888957360123608\\
50	1.33267800608452\\
100	1.6063075\\
150	1.74427083333333\\
200	2.06552623166667\\
250	2.18930209166667\\
};

\node[draw,align=left,scale=0.75,anchor=south east] at (axis cs: 250,0.8) {Solid: multi-slope profile\\Dashed: constant profile};

\end{axis}
\end{tikzpicture}%
\caption{Average sum-rate in the multi-UE MISO scenario under the proposed RIS design.}
\label{MISO}
\vspace{-0.5cm}
\end{figure}

Next we present the simulation result for the multi-UE MISO case with $Q = 8$, $\Delta_{T} = 7.5 \lambda_{0}$, $P = 10$~mW, $N_{s} = 256$, $N_{y} = 200$, $N_{z} = 1$, and $b = 4$. We consider the UEs to be distributed on a half circle facing the RIS at a distance of $r = 15$~m. The channels between the Tx and RIS, as well as between the RIS and UEs, are assumed to undergo Rician fading with $\kappa = 20$.  Fig. \ref{MISO} depicts the average sum rate for $L = 3$ and $L = 6$ UEs while changing $N_{y}$ for the two considered precoding schemes, i.e.  MRT and ZF. Notably, the $\arctan$ phase-frequency profile significantly outperforms the constant phase-frequency profile under MRT precoding (a $43 \%$ improvement in the achievable over the constant phase-frequency profile is observed at $N_{y} = 250$), while there is no noticeable gain under ZF precoding. This can be explained by the fact that  MRT maximizes the desired signal power, which coincides with the objective used to select the sets $\mathcal{M}$ and $\mathcal{I}_{\tilde{m}}$, while ZF minimizes the interference, which reduces the power gain of using the $\arctan$ phase-frequency profile model. An interesting future direction is to design the sets based on an optimal solution that maximizes the SINR and not just the received signal power as done in \eqref{eq25}. We also observe that increasing $N$ results in a larger average sum-rate gain under the $\arctan$ phase-frequency model over the constant phase-frequency model. 

%We note that the gain of $\arctan$ phase-frequency profile over the constant one in the case of MRT precoding is smaller than the maximum gain shown in table \ref{NsubcarrierTable}, because we are distributing the users over random locations .i.e., near specular the constant phase-frequency profile provides better response than the $\arctan$ profile while for other user location the $\arctan$ profile performs better. 

\vspace{-0.5cm}
\section{Conclusion}\label{conclusion}

We considered a practical RIS model and analyzed the amplitude- and phase-frequency profiles of the reflection coefficient of each RIS unit cell. Different from existing works that assume a frequency-flat phase response for the RIS unit cells, we provided a simple mathematical model to approximate the phase-frequency relationship that is parameterized by the slope and shift-shift of the phase variation over frequency. These parameters can take values from discrete sets, which are selected using an algorithm we proposed for an RIS-assisted single-user wide-band OFDM setting. Specifically, the algorithm selects the sets of slopes and shifts that the proposed RIS design should provide to maximize the received signal strength over a given geographical area. We then outlined a low-complexity algorithm to jointly optimize the power allocations across the sub-carriers and the phase-frequency profiles imparted by the RIS unit cells as constrained by the designed sets to maximize the rate. The proposed algorithms were also extended to the multi-user MISO scenario. Numerical results highlighted the importance of accounting for phase-frequency dependency in designing and analyzing RIS-assisted wide-band systems. Specifically, up to $91 \%$ improvement in the achievable rate was realized with the proposed model in the single-UE SISO wide-band OFDM system with large RISs, and up to $43 \%$ improvement was realized in the multi-UE MISO system with MRT precoding. Future research directions include improving the phase-frequency profiles selection and optimization algorithms in different scenarios, such as the multi-UE MISO setting with ZF precoding.

\bibliographystyle{IEEEtran}
\bibliography{IEEEabrv,bibliography}

\begin{thebibliography}{10}
\providecommand{\url}[1]{#1}
\csname url@samestyle\endcsname
\providecommand{\newblock}{\relax}
\providecommand{\bibinfo}[2]{#2}
\providecommand{\BIBentrySTDinterwordspacing}{\spaceskip=0pt\relax}
\providecommand{\BIBentryALTinterwordstretchfactor}{4}
\providecommand{\BIBentryALTinterwordspacing}{\spaceskip=\fontdimen2\font plus
\BIBentryALTinterwordstretchfactor\fontdimen3\font minus
  \fontdimen4\font\relax}
\providecommand{\BIBforeignlanguage}[2]{{%
\expandafter\ifx\csname l@#1\endcsname\relax
\typeout{** WARNING: IEEEtran.bst: No hyphenation pattern has been}%
\typeout{** loaded for the language `#1'. Using the pattern for}%
\typeout{** the default language instead.}%
\else
\language=\csname l@#1\endcsname
\fi
#2}}
\providecommand{\BIBdecl}{\relax}
\BIBdecl

\bibitem{mmwave}
J.~Karjalainen \emph{et~al.}, ``Challenges and opportunities of mm-wave
  communication in {5G} networks,'' in \emph{Int. Conf. on Cogn. Radio Oriented
  Wireless Networks and Commun.}, 2014, pp. 372--376.

\bibitem{MIMOCom}
S.~Mohanty \emph{et~al.}, ``Design and {BER} performance analysis of {MIMO} and
  massive {MIMO} networks under perfect and imperfect {CSI},'' in \emph{Fourth
  Int. Conf. on I-SMAC}, 2020, pp. 307--312.

\bibitem{Cluster}
Y.~Liu \emph{et~al.}, ``Grouping and cooperating among access points in
  user-centric ultra-dense networks with non-orthogonal multiple access,''
  \emph{IEEE J. on Sel. Areas in Commun.}, vol.~35, no.~10, pp. 2295--2311,
  2017.

\bibitem{usercentric}
J.~Liu and H.~Zhang, ``Power allocation in ultra-dense networks through deep
  deterministic policy gradient,'' \emph{IEEE Wireless Commun. Letters},
  vol.~11, no.~12, pp. 2502--2506, 2022.

\bibitem{noma}
A.~S. Marcano and H.~L. Christiansen, ``Performance of non-orthogonal multiple
  access ({NOMA}) in mmwave wireless communications for {5G} networks,'' in
  \emph{2017 Int. Conf. on Comput., Netw. and Commun. (ICNC)}, 2017, pp.
  969--974.

\bibitem{machinedense}
W.-S. Liao \emph{et~al.}, ``Machine learning-based signal detection for {CoMP}
  downlink in ultra-dense small cell networks,'' \emph{IEEE Access}, vol.~8,
  pp. 17\,454--17\,463, 2020.

\bibitem{Relay}
S.~P. Padhy \emph{et~al.}, ``Performance evaluation of relays used for next
  generation wireless communication networks,'' in \emph{2018 Int. Conf. on
  Appl. Electromagnetics, Signal Process. and Communication (AESPC)}, vol.~1,
  2018, pp. 1--4.

\bibitem{RISoverview}
M.~Jian \emph{et~al.}, ``Reconfigurable intelligent surfaces for wireless
  communications: Overview of hardware designs, channel models, and estimation
  techniques,'' \emph{Intell. and Converged Networks}, vol.~3, no.~1, pp.
  1--32, 2022.

\bibitem{RISPro}
R.~Fara \emph{et~al.}, ``A prototype of reconfigurable intelligent surface with
  continuous control of the reflection phase,'' \emph{IEEE Wireless Commun.},
  vol.~29, no.~1, pp. 70--77, 2022.

\bibitem{RISPolar}
Y.~Zhang \emph{et~al.}, ``A new sandwich linear polarization and frequency
  selective surface design,'' in \emph{Proc. of 2014 3rd Asia-Pacific Conf. on
  Antennas and Propagation}, 2014, pp. 940--942.

\bibitem{THzBlockedLoS}
H.~Du \emph{et~al.}, ``Performance and optimization of reconfigurable
  intelligent surface aided {THz} communications,'' \emph{IEEE Trans. on
  Commun.}, vol.~70, no.~5, pp. 3575--3593, 2022.

\bibitem{MISORIS}
Q.-U.-A. Nadeem \emph{et~al.}, ``Asymptotic max-min {SINR} analysis of
  reconfigurable intelligent surface assisted {MISO} systems,'' \emph{IEEE
  Trans. on Wireless Commun.}, vol.~19, no.~12, pp. 7748--7764, 2020.

\bibitem{MIMORIS}
K.~Ardah \emph{et~al.}, ``Double-{RIS} versus single-{RIS} aided systems:
  Tensor-based {MIMO} channel estimation and design perspectives,'' in
  \emph{ICASSP 2022 - 2022 IEEE Int. Conf. on Acoust., Speech and Signal
  Process. (ICASSP)}, 2022, pp. 5183--5187.

\bibitem{NOMARIS}
B.~Zhao \emph{et~al.}, ``Ergodic rate analysis of {STAR-RIS} aided {NOMA}
  systems,'' \emph{IEEE Commun. Letters}, vol.~26, no.~10, pp. 2297--2301,
  2022.

\bibitem{IRSOFDM}
W.~Jiang \emph{et~al.}, ``Joint transmit precoding and reflect beamforming for
  {IRS}-assisted {MIMO}-{OFDM} secure communications,'' in \emph{2021 IEEE
  Global Commun. Conf. (GLOBECOM)}, 2021, pp. 1--6.

\bibitem{RISstableband}
G.~C. {Alexandropoulos} \emph{et~al.}, ``{{RIS}-Enabled Smart Wireless
  Environments: Deployment Scenarios, Network Architecture, Bandwidth and Area
  of Influence},'' \emph{arXiv e-prints}, p. arXiv:2303.08505, Mar. 2023.

\bibitem{PracPin}
S.~Hassouna \emph{et~al.}, ``Discrete phase shifts for intelligent reflecting
  surfaces in {OFDM} communications,'' in \emph{2022 Int. Workshop on Antenna
  Technol. (iWAT)}, 2022, pp. 128--131.

\bibitem{PracPhaseAmp2}
S.~Abeywickrama, R.~Zhang, and C.~Yuen, ``Intelligent reflecting surface:
  Practical phase shift model and beamforming optimization,'' in \emph{ICC 2020
  - 2020 IEEE Int. Conf. on Commun. (ICC)}, 2020, pp. 1--6.

\bibitem{PracPhaseAmp}
H.~Li \emph{et~al.}, ``Intelligent reflecting surface enhanced wideband
  {MIMO}-{OFDM} communications: From practical model to reflection
  optimization,'' \emph{IEEE Trans. on Commun.}, vol.~69, no.~7, pp.
  4807--4820, 2021.

\bibitem{RISPin}
W.~Xue \emph{et~al.}, ``Design of 1-{B}it digital coding reconfigurable
  reflectarray using aperture-coupled elements controlled by {PIN} diodes,'' in
  \emph{2018 Int. Conf. on Microwave and Millimeter Wave Technology (ICMMT)},
  2018, pp. 1--3.

\bibitem{RISRing}
C.~Liu \emph{et~al.}, ``Design of an {E-B}and 1-bit reconfigurable reflectarray
  antenna using {PIN} diodes,'' in \emph{2022 16th European Conf. on Antennas
  and Propagation (EuCAP)}, 2022, pp. 1--3.

\bibitem{PlannerWave}
A.~Amar and A.~Weiss, ``Direct position determination of multiple radio
  signals,'' in \emph{2004 IEEE Int. Conf. on Acoust., Speech, and Signal
  Process.}, vol.~2, 2004, pp. ii--81.

\bibitem{RicianChannel}
J.~Abraham \emph{et~al.}, ``Statistics of the effective massive {MIMO} channel
  in correlated {Rician} fading,'' \emph{IEEE Open J. of Antennas and
  Propagation}, vol.~3, pp. 238--248, 2022.

\bibitem{ZeroForcing}
G.~Liu \emph{et~al.}, ``Outage analysis for {MISO} zero-forcing precoding with
  outdated {CSI},'' \emph{IEEE Trans. on Veh. Technol.}, vol.~69, no.~8, pp.
  9152--9156, 2020.

\bibitem{MaxRatTRan}
T.~Lo, ``Maximum ratio transmission,'' \emph{IEEE Trans. on Commun.}, vol.~47,
  no.~10, pp. 1458--1461, 1999.

\end{thebibliography}

\end{document}